\documentclass[preprint,12pt]{elsarticle}

\usepackage{amssymb}
\usepackage{subcaption,graphicx}
\usepackage{caption}
\usepackage{amsmath}
\usepackage{bm}
\usepackage[version=4]{mhchem}
\usepackage{lineno}

\newcommand{\de}{\mbox{d}}
\newcommand{\ghat}{\widehat{\dot\gamma}}
\newcommand{\Ghat}{\widehat{\Gamma}}

\usepackage{xcolor}

\begin{document}

\begin{frontmatter}



\title{Micromechanics of liquid-phase exfoliation of a layered 2D material: a hydrodynamic peeling model}


\author[qmul]{Giulia Salussolia}
\author[jamstec]{Ettore Barbieri}
\author[trento,qmul,ket]{Nicola Maria Pugno}
\author[qmul, delft]{Lorenzo Botto}
\ead{l.botto@qmul.ac.uk}
\cortext[cor1]{Corresponding author}

\address[qmul]{School of Engineering and Materials Science, Queen Mary University of London,\\ Mile End Road, E1 4NS, London, UK}

\address[delft]{Process and Energy Department, TU Delft, Leeghwaterstraat, 2628 CB, Delft, The Netherlands}

\address[jamstec]{Japan Agency for Marine-Earth Science and Technology (JAMSTEC)\\
Research Institute for Value-Added-Information Generation (VAiG)\\
Center for Mathematical Science and Advanced Technology (MAT)\\
3173-25, Showa-machi, Kanazawa-ku, Yokohama-city, Kanagawa, 236-0001, Japan
}
\address[trento]{Laboratory of Bio-Inspired \& Graphene Nanomechanics, Department of Civil, Environmental and Mechanical Engineering, University of Trento, Trento, Italy}
\address[ket]{KET Labs, Edoardo Amaldi Foundation, Rome, Italy}

\begin{abstract}
We present a  micromechanical analysis of flow-induced peeling of a layered 2D material suspended in a liquid, for the first time accounting for realistic hydrodynamic loads. In our  model, fluid forces trigger a fracture of the inter-layer interface by lifting a flexible ``flap'' of nanomaterial from the surface of a suspended microparticle.  We show that the so far ignored dependence of the hydrodynamic load on the wedge angle produces a transition in the curve relating the critical fluid shear rate for peeling to the non-dimensional adhesion energy. For intermediate values of the non-dimensional adhesion energy, the critical shear rate saturates, yielding critical shear rate values that are drastically smaller than those predicted by a constant load assumption. Our results highlight the importance of accounting for realistic hydrodynamic loads in fracture mechanics models of liquid-phase exfoliation. 
\end{abstract}




\end{frontmatter}


\section{INTRODUCTION}
\label{sec:intro}
Atomically thin, two-dimensional materials such as graphene, boron nitride, or  \ce{MoS2} have attracted enormous interest recently \citep{miro2014}. As a consequence the nanomechanics of 2D nanomaterials has emerged as an important direction in the solid mechanics literature. Much of the work on the mechanics of 2D nanomaterials has focused on solid mechanics and tribological aspects, such as adhesion \cite{bunch2012adhesion}, tearing \citep{sen2010tearing}, scrolling \citep{shi2010tunable,shi2010mechanics}, buckling \citep{jiang2014interfacial,chen2012plane}, wrinkling \citep{zhang2014understanding,zang2013multifunctionality} and friction \citep{pugno2013generalization}. Of interest is typically the deformation of the solid structure. However, the effect of the medium in which the solid structure is immersed is often not considered, particularly when the medium is a fluid. Many wet processes involving 2D materials and mechanical tests of 2D materials  are carried out in liquids or in contact with liquids \citep{van2017, miskin2017, chen2012vortex, arao2016,akbari2016}. When liquids are present, not only the interfacial thermodynamics changes \citep{stoloff1963,johnson1971}. One has also to consider the possible coupling to flow. 

Several techniques have been developed to produce 2D nanomaterials: bottom-up methods such as chemical vapour deposition \citep{mattevi2011review} and epitaxial growth \citep{yang2013epitaxial} are used to build layers of material from its molecular components, while in top-down methods the layers of a bulk multilayer particle are separated through electrochemical exfoliation \citep{abdelkader2015get}, ball-milling \citep{jeon2012large} or liquid-assisted processes like sonication \citep{hernandez2008high} and shear mixing \citep{paton2014scalable}. In the current paper, we focus on mechanical aspects of liquid-phase exfoliation by shear mixing, a scalable process to produce 2D nanomaterials on industrial scales (for a comprehensive review, see \citep{backes2016}). In liquid-phase exfoliation, plate-like microparticles of layered materials are suspended in a liquid solvent. The liquid is then mixed energetically under turbulent conditions \citep{varrla2014}. Each microparticle is formed by stacks of hundreds or thousands of atomic layers, bound together by relatively weak inter-layer forces of the van der Waals type. If the solvent is chosen appropriately and the intensity of the turbulence sufficiently high, the large fluid dynamic forces applied to each suspended particle can overcome interlayer adhesion, ultimately producing single- or few-layer nanosheets. Choosing the optimal shear intensity level is paramount, as too small hydrodynamic forces will not induce exfoliation while too large forces will damage each sheet. An understanding of how the exfoliation process occurs at the microscopic level is currently lacking. 

A recent exfoliation model based on a sliding deformation has been proposed by Paton et al. \citep{paton2014scalable}, as an extension of a previous work by Chen et al. \citep{chen2012vortex}. In this model, the force for sliding is calculated by considering the change in adhesion energy (accounting for changes in liquid-solid, solid-solid and liquid-liquid surface areas) as the overlap between two 2D nano-layers changes. Paton's model predicts a critical shear rate proportional to the adhesion energy, and inversely proportional to the first power of the platelet lateral dimension. Because in the model the sheets are considered infinitely rigid, the results are independent of the mechanical properties of the sheets. For instance, the bending rigidity of the sheets does not appear in the expression for the critical shear rate. Since the seminal scotch-tape experiment of Geim and Novoselov \citep{novoselov2004}, it has become clear that the interplay of solid deformation and adhesion  can play a fundamental role in triggering layer detachment. 

A simple physical model that is sensitive to mechanics is to assume that the effect of the fluid is to peel off layers of nanomaterial, inducing a fracture of the van der Waals interface. In addition to being  physically plausible, such model would explain why in liquid-phase exfoliation removal of layers occurs first at the outer surface of a mother particle \citep{alaferdov2014}. Flow-induced fracture phenomena have been studied extensively for colloidal aggregates composed of roughly spherical  beads \citep{kendall1988, de2014}. Models for exfoliation of plate-like, layered particles due to microscopic peeling have been proposed in the context of clay-filled polymer nanocomposites \citep{borse2009}\citep{cho2004}, but have not been rigorously justified from the point of view of the coupling between flow and deformation mechanics. For instance, these models rely on strong, often irrealistic assumptions regarding the hydrodynamic load distribution. For example, in the model of Ref. \citep{borse2009}  the fluid  is assumed to exert a constant, tangential force on the peeled layer  at a specified angle. In reality, one should expect a dependence of the flow-induced forces on the configuration of the peeled layer and that normal forces will play an important role. The consequence of such dependence is so far unknown.  The forces required to detach a layer in a peeling problem depend strongly on the peeling angle \citep{kendall1975}. But, in a peeling problem in which the flow produces the load, the peeling angle cannot be controlled independently, as this quantity depends on the deformation of the solid structure. In addition, hydrodynamic forces are distributed over the whole surface of the peeled layer, including the edge. In fracture problems, different assumptions regarding the load distribution can result in different predictions even if the total force values at play are the same \citep{capozza2012}. Considering realistic hydrodynamic loads is crucial to develop exfoliation models that will withstand future experimental tests.  

\begin{figure}[htbp]
\centering
\includegraphics[width=0.59\textwidth]{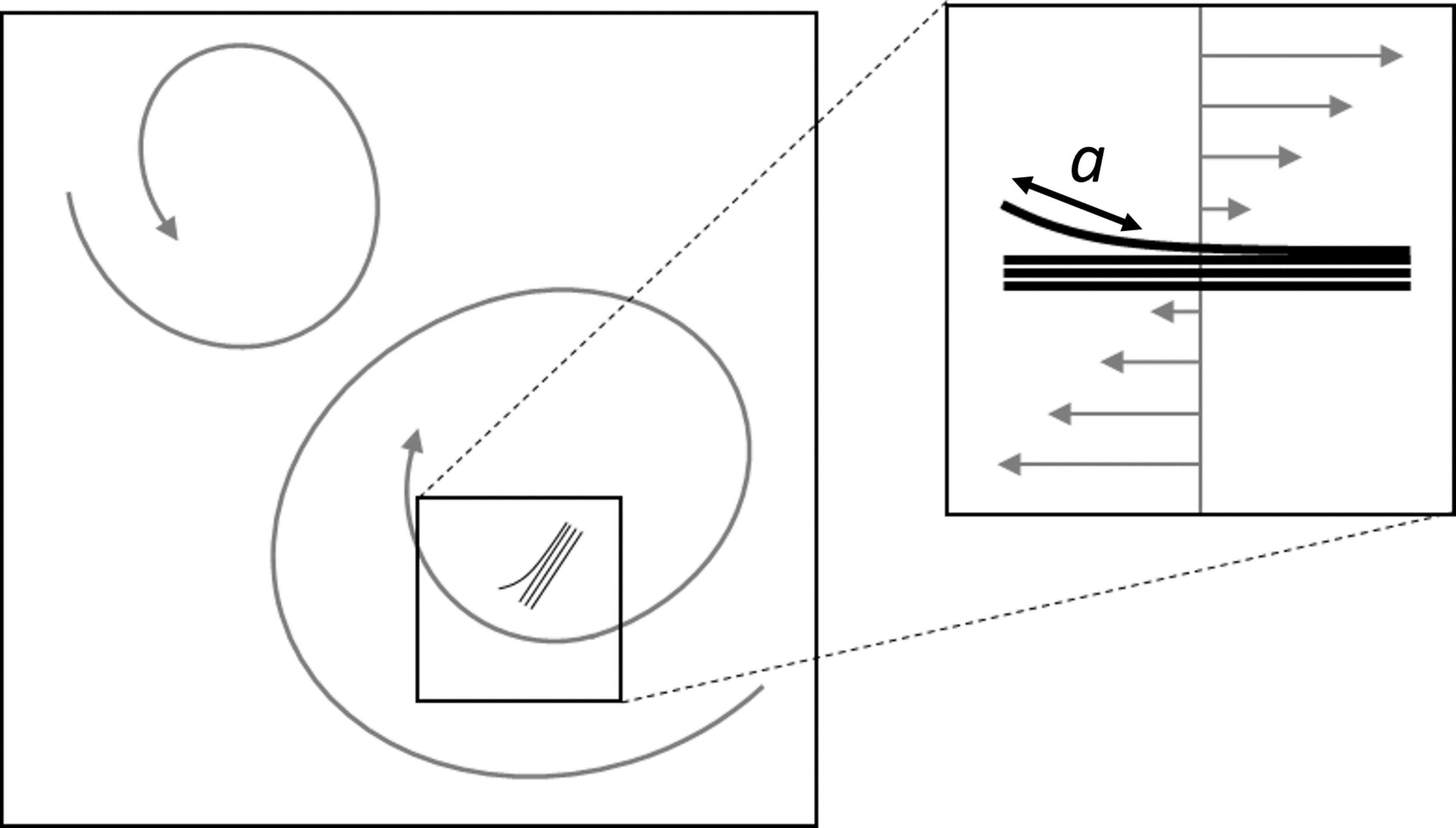}
\caption{(a) A layered 2D material microplate suspended in a turbulent flow.  The ambient flow in the neighbourhood of the particle can be approximated as a simple shear flow.   }
\label{fig:schematic}
\end{figure}

In our paper, we analyse a ``hydrodynamic peeling'' model of exfoliation, not making strong assumptions regarding the magnitude and distribution of the hydrodynamic load. Rather, we calculate the load from first principles, using high-resolution simulations of the Stokes equations for a simplified geometry. The forces computed from these high-fidelity simulations are then used to calculate the deformation of the solid. Griffith's theory for crack initiation is then used  to quantify the critical (fluid) shear rate for exfoliation as a function of the relevant adhesion, mechanical and geometrical parameters of the suspended particle (Fig. \ref{fig:schematic}b). When possible, we derive explicit formulas for the critical shear rate. This quantity is essential to predict the kinetics of exfoliation  \citep{babler2012} and the operating parameters for ``optimal'' exfoliation. In the model, the flap is approximated as a continuum sheet. A continuum representation is justified by the good separation of scale between the length of each nanosheet (typically in the micron range) and the characteristic length of the nanostructural elements (e.g. the period of the crystal lattice in graphene).

\section{PROBLEM DESCRIPTION}
\label{sec:problem}

Consider a microplate of layered 2D material (e.g. a graphite microplate \citep{wick2014}) suspended in a turbulent flow. In correspondence to one of the layers the inter-layer interface presents an initial flaw of length $a$, where the molecular bonds are already broken. A ``flap'' of length $a$ forms which is detached from the mother particle.  We are interested in relating the critical fluid shear rate for interfacial crack initiation to the bending rigidity of the flap, the inter-layer adhesion energy, and the flap size. 

In our analysis we assume that $a>0$. Our results are relevant to the case in which a debonding of the interface had already occurred in the proximity of the edge, for instance due to molecular intercalation by the surrounding liquid. Borse and Kamal made a similar assumption in the context of clay exfoliation in polymers \citep{borse2009}. 

If the lateral size of the microplate is smaller than the smallest turbulent flow scales, the instantaneous ambient flow ``seen'' by the particle can be approximated as a locally Stokes flow, characterised by different degrees of extension, shear and rotation depending on the position and orientation of the particle \citep{biferale2014}. Purely rotational contributions to the ambient flow induce a rotation of the particle, but no significant net load on the flap. Purely extensional contributions are important if the  shift between the layers is large (i.e. the layers are not ``in registry''), a situation that we do not consider here. As a consequence, the local ambient flow can be approximated, to leading order, as a simple shear flow (Fig. \ref{fgr:flap}).

The question is: what is the load distribution corresponding to this shear flow? The hydrodynamic force distribution on a particle suspended in a shear flow and presenting a flap has not been studied so far (we only found work on hydrodynamic forces on rigid fences attached to solid walls \citep{jeong1983slow,taneda1979visualization,higdon1985stokes}). To quantify the hydrodynamic load on the flap, in Section \ref{sec:flow} we therefore propose a fluid dynamics analysis based on high-resolution flow simulations of a simplified flap geometry. In the flow simulations the particle is exposed to a simple shear flow of strength $\dot{\gamma}$. Jeffery's theory for the rotational dynamics of for plate-like particles predicts that a particle of aspect ratio $\Lambda$ rotates in a shear flow, but spends a  time of the order of $\Lambda \dot{\gamma}^{-1}$ oriented with the flow \cite{jeffery1922motion}. Microparticles of 2D materials tend to have very large aspect ratios ($\Lambda \sim 1000$ \citep{wick2014}). Hence, in our fluid dynamics simulations we will assume that the long axis of the particle is aligned with the undisturbed streamlines of the shear flow field. 

The load extracted from the simulation will be fitted to analytical functions. In Sec. \ref{sec:analytical} such functions are used in a solid mechanics model  to predict the critical shear rate for exfoliation $\dot{\gamma}_c$, obtained from Griffith's theory assuming brittle fracture \citep{griffith1921}.

\section{RESULTS}
\subsection{Analysis of the hydrodynamic load}
\label{sec:flow}

\begin{figure}[htbp]
\centering
\includegraphics[width=0.9\textwidth]{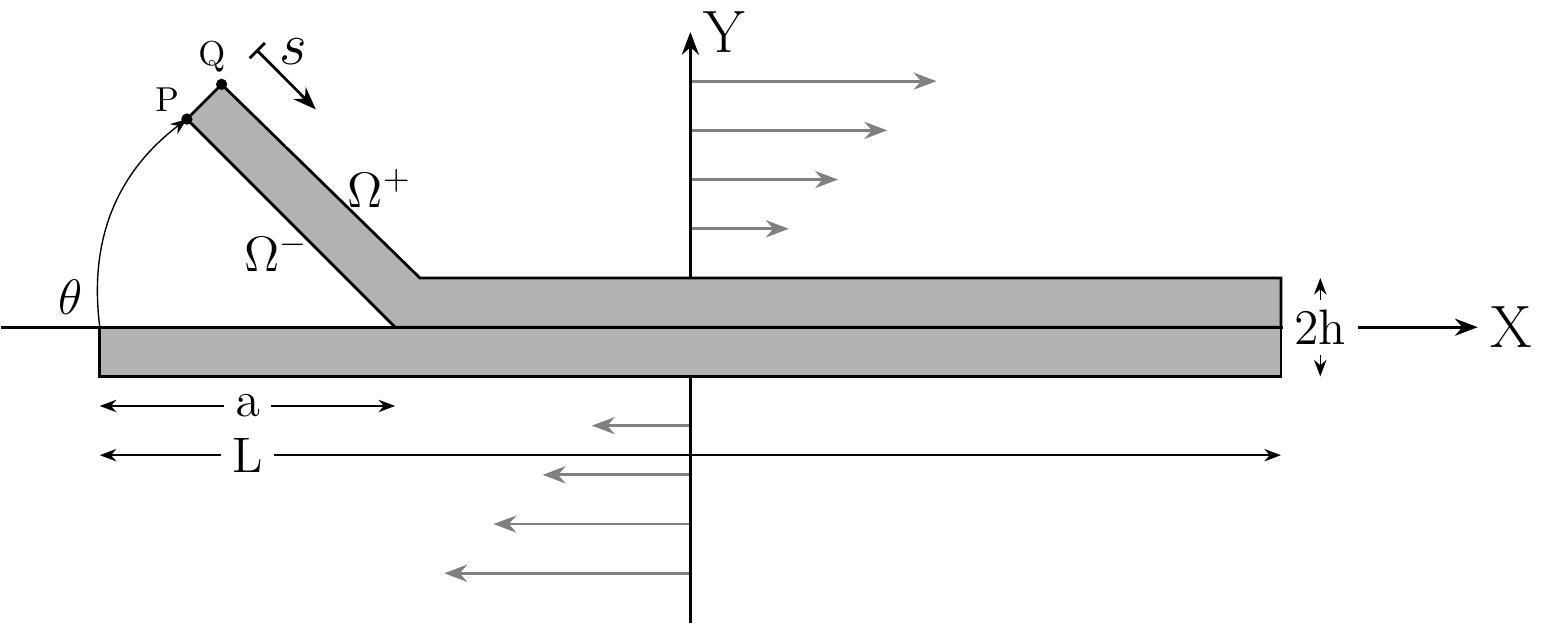}
\caption{Simplified geometry used in the flow simulation. }
\label{fgr:flap}
\end{figure}

The simplified geometry for the flow simulation is presented in Fig.  \ref{fgr:flap}. In this configuration, the flap is straight and the wedge is parametrised by the flap length $a$ (also equal to the length of the initial flaw) and the wedge opening angle $\theta$. We will see that the pressure in the wedge is approximately constant. So, neglecting the flap curvature in the calculation of the hydrodynamic load does not induce a large error.   The surface of the flap is composed of three surfaces: the lower surface $\Omega^-$ in contact with the fluid in the wedge, the upper surface $\Omega^+$ exposed to the outer flow, and the edge surface $\Omega^e$ between the corner points P and Q. A coordinate $s$  running from the edge of the flap (point P) to the point of intersection of the flap with the horizontal plate will be used to discuss the hydrodynamic stress profiles.  A coordinate $s_e$, running from the points P and Q, will be used to discuss the hydrodynamic stress profile along the edge . The bottom layer and the flap  have the same thickness, $h$, and length, $L$. In the flow simulations we kept $L$ and  $a$ fixed and changed $h$ and $\theta$. We sought results that are independent of $h$ by examining simulations for decreasing values of this parameter.

The simulations were carried out with the commercial software ANSYS FLUENT. We solved the incompressible Stokes equations (corresponding to a negligible particle Reynolds number) in a rectangular domain $[-X_D,X_D]\times[-Y_D,Y_D]$ surrounding the particle. Periodic boundary conditions were enforced at the boundaries $X=-X_D$ and $X=X_D$ ( $X=0$ corresponding to the particle centre). At the boundaries $Y=-Y_D$ and $Y=Y_D$, we prescribe a tangential velocity $u=\pm \dot{\gamma} Y_D$ and zero normal velocity. No-slip is assumed at the particle surface.
The computational mesh used is non-uniform. A triangular mesh is used in the wedge region and a structured quadrilateral mesh is used in the rest of the domain.  To ensure adequate resolution, the typical mesh size $\Delta X$ is much smaller than the thickness $h$ of each layer (we typically use   $\Delta X\approx 0.1\;h$ in the flap edge region, and much smaller values of $\Delta X$ around the points P and Q). 

In principle, the solution of the fluid mechanics problem is coupled with the solution of the solid mechanics problem providing the deformation of the flap. Solving the two-way coupled problem numerically is possible, for example by using iterations \citep{pozrikidis2011}. However, the advantage of the one-way coupled approach we adopt is that  explicit analytical expressions relating the critical fluid shear rate to the relevant geometric and mechanical variables can be obtained. The hydrodynamic load for a straight flap and for a curved flap are expected to be similar. We will see that the fluid pressure within the wedge is approximately constant. As a result the normal force on the flap expected to depend primarily on the aperture angle and flap length, and only marginally on the details of the flap shape.

\begin{figure}[htb]
\centering
\begin{subfigure}{0.8\textwidth}
\includegraphics[width=\textwidth]{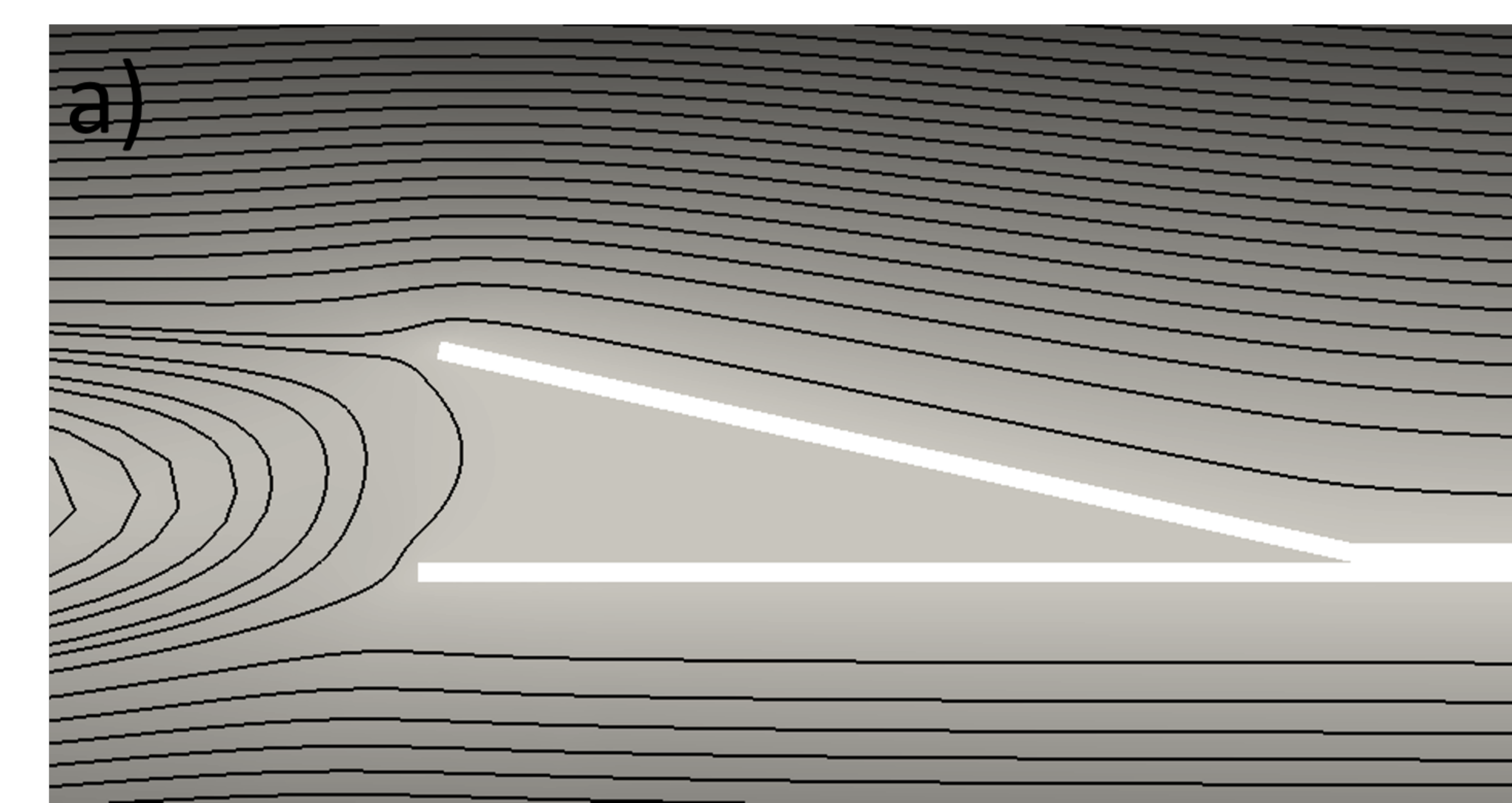}
\end{subfigure}
\begin{subfigure}{0.8\textwidth}
\includegraphics[width=\textwidth]{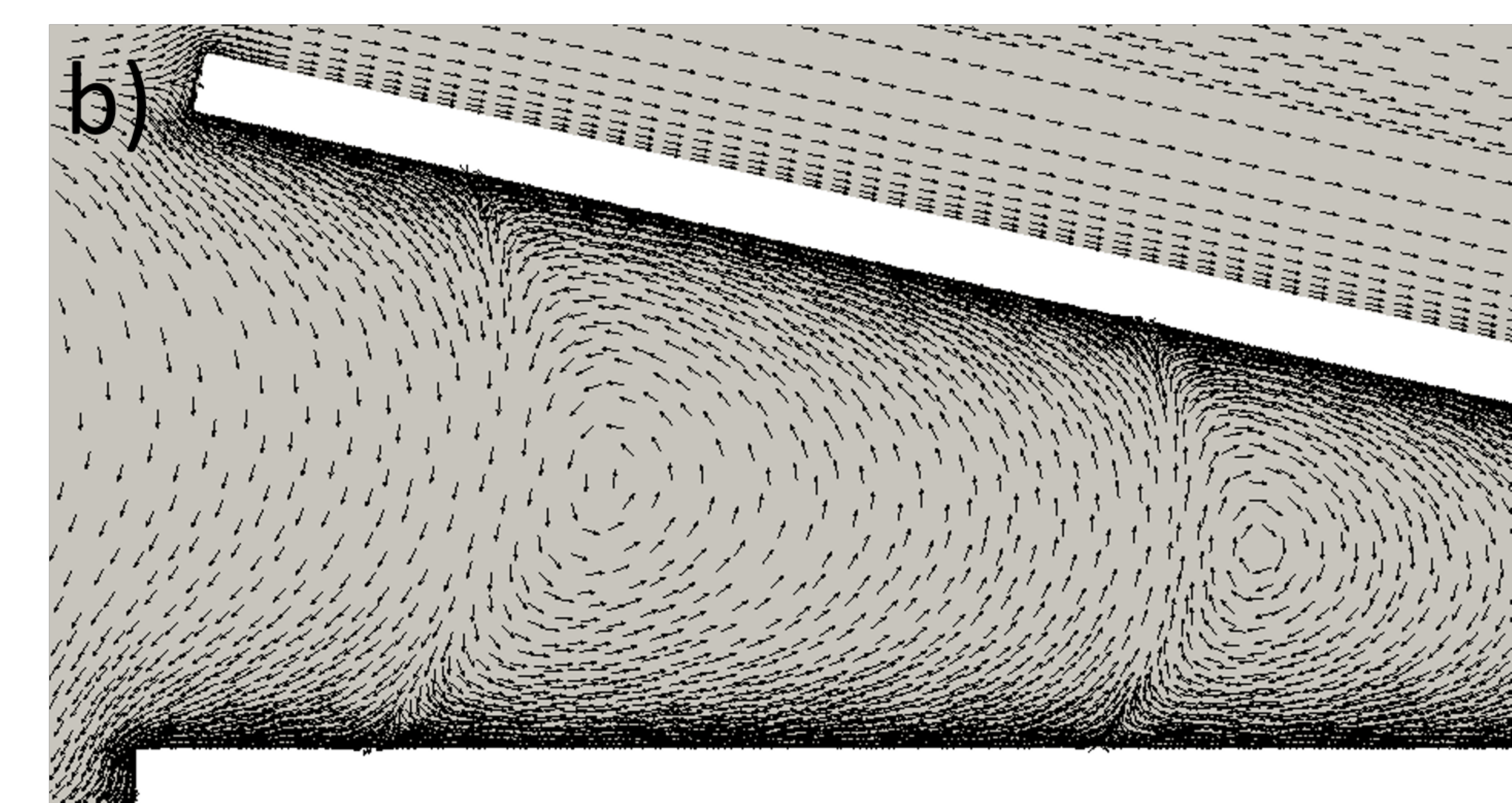}
\end{subfigure}
\caption{(a) Streamlines around the particle. (b) Detail of the recirculating eddies in the wedge below the flap.}
\label{fgr:velocity}
\end{figure}

The general features of the flow around the model particle are illustrated in Fig. \ref{fgr:velocity}a and \ref{fgr:velocity}b. For small values of $\theta$, the streamlines run almost parallel to the exterior surfaces of the particle. The streamlines need to curve sharply near the entrance of the wedge. As a consequence, a sequence of counter-rotating eddies form in the wedge region (Fig. \ref{fgr:velocity}b). The characteristic velocity in these eddies decays very fast as the wedge tip is approached \citep{moffatt1964viscous}. Hence, the fluid in the wedge can be consider practically quiescent in comparison to the fluid regions outside of the wedge (where velocities are of the order of $\dot\gamma a \theta$). An important consequence of this observation is that  the pressure in the wedge region is approximately uniform. In the region near the edge, on the other hand, velocity gradients are large and the pressure variation is considerable. 

\begin{figure}
\noindent
\begin{minipage}[t]{0.49\textwidth}
\raggedleft
\includegraphics[width=\textwidth]{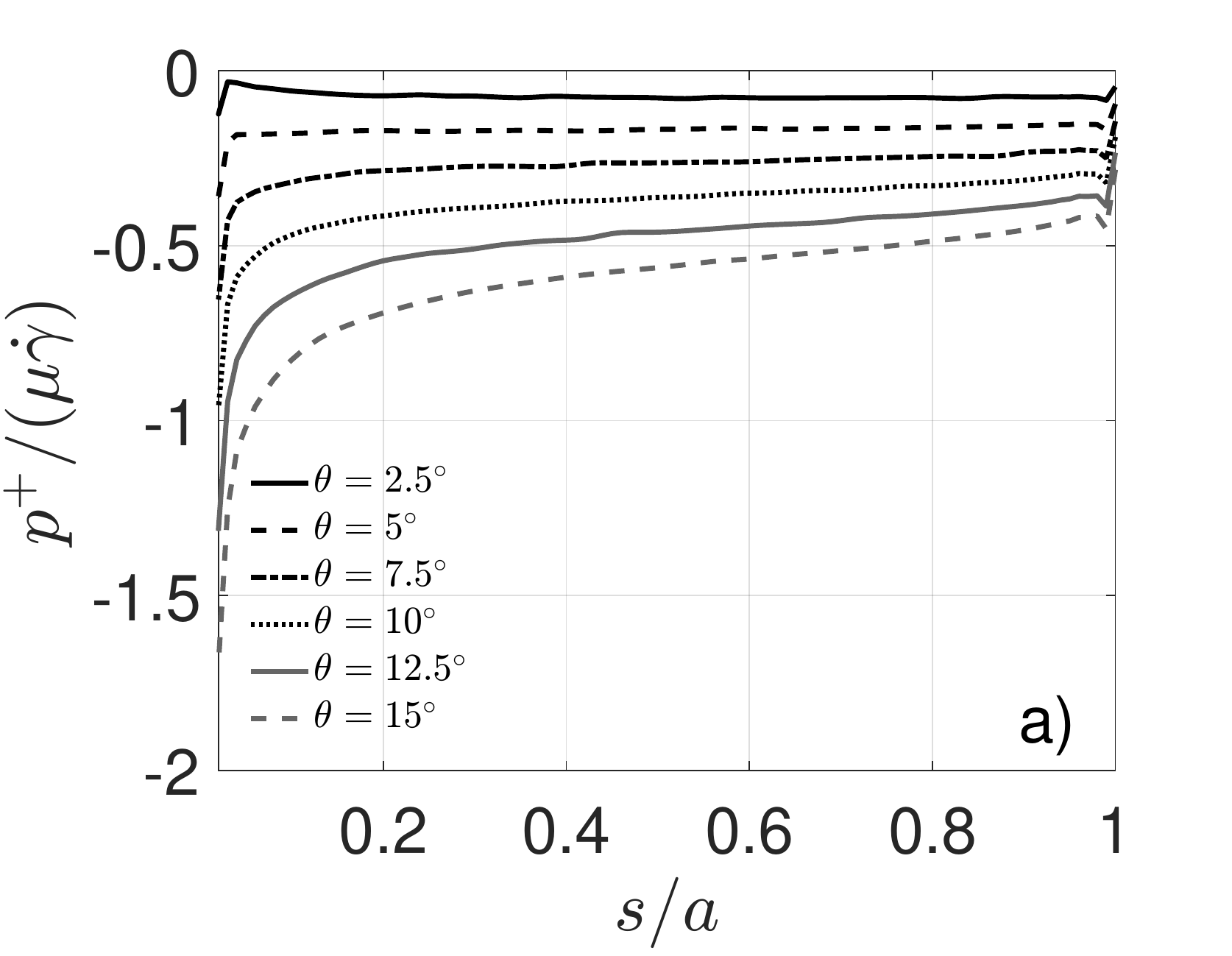}
\includegraphics[width=\textwidth]{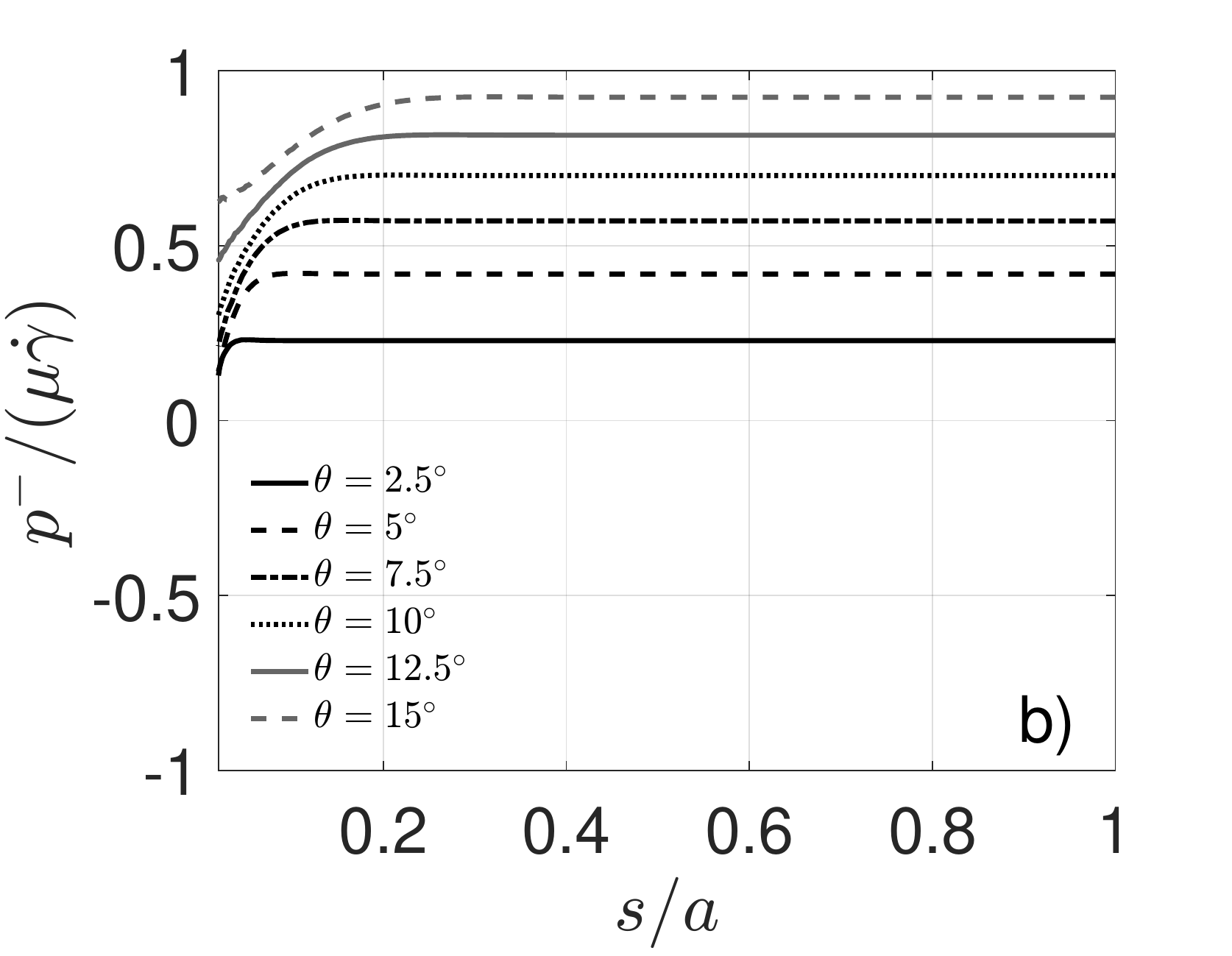}
\includegraphics[width=\textwidth]{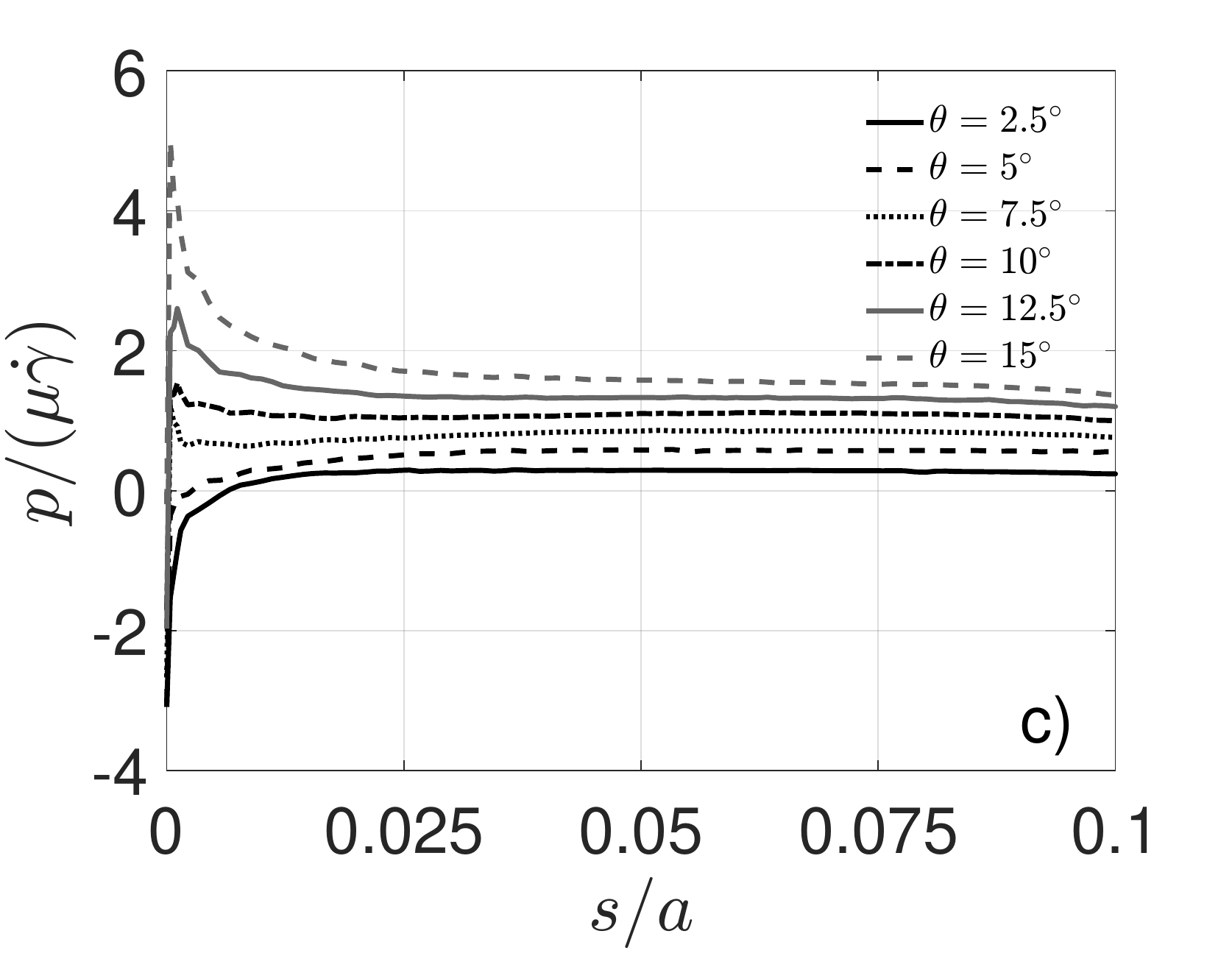}
\caption{(a) and (b): Pressure distribution on the surfaces $\Omega^+$ and $\Omega^-$, respectively. The thickness of each layer is $h= a/50$. (c) Total pressure $p=p^--p^+$ in the region $0 \leq s/a\leq 0.1$ near the edge of the flap.}
\label{fgr:pressuredistribution}
\end{minipage}%
\hfill
\begin{minipage}[t]{0.49\textwidth}
\raggedright
\includegraphics[width=\textwidth]{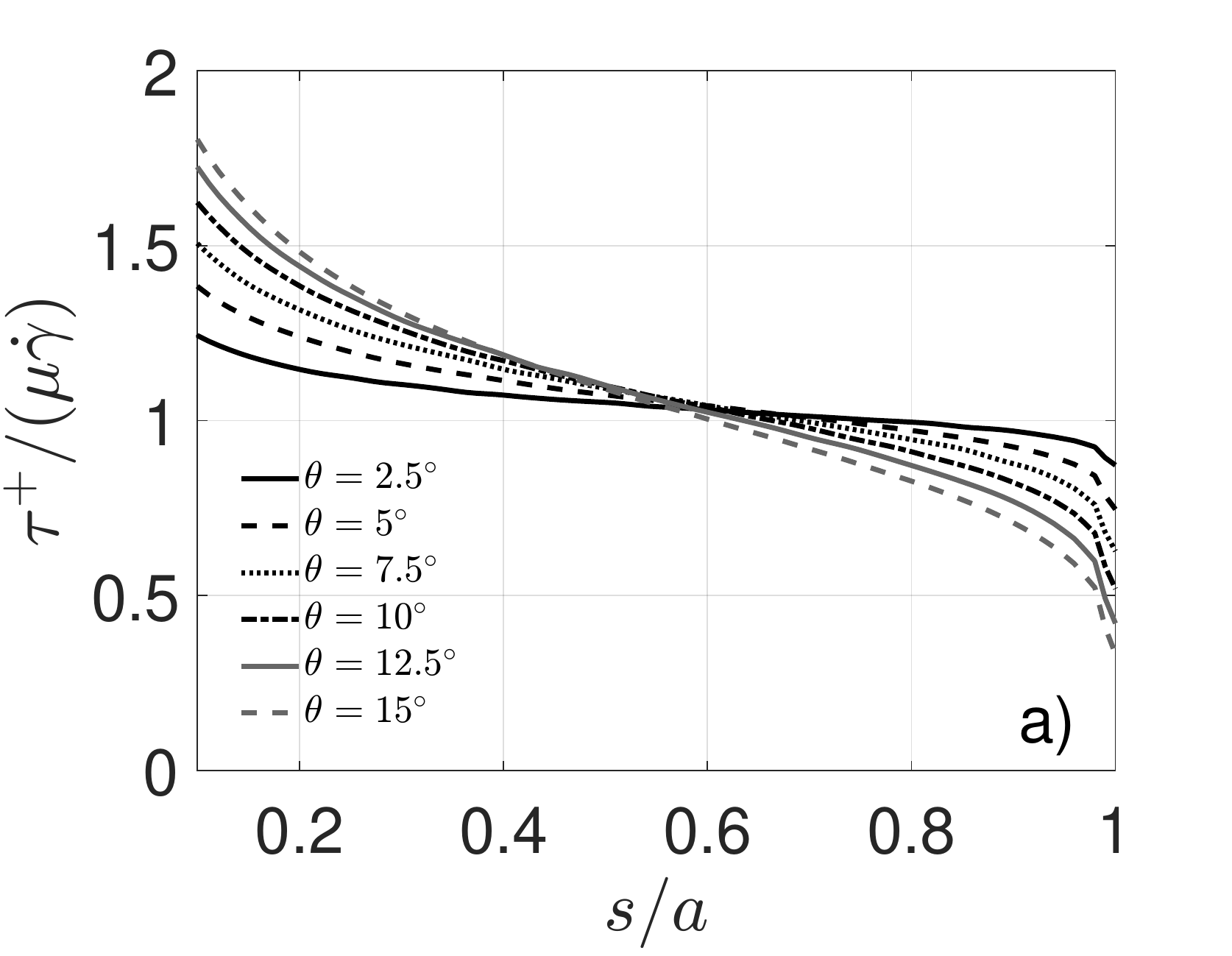}
\includegraphics[width=\textwidth]{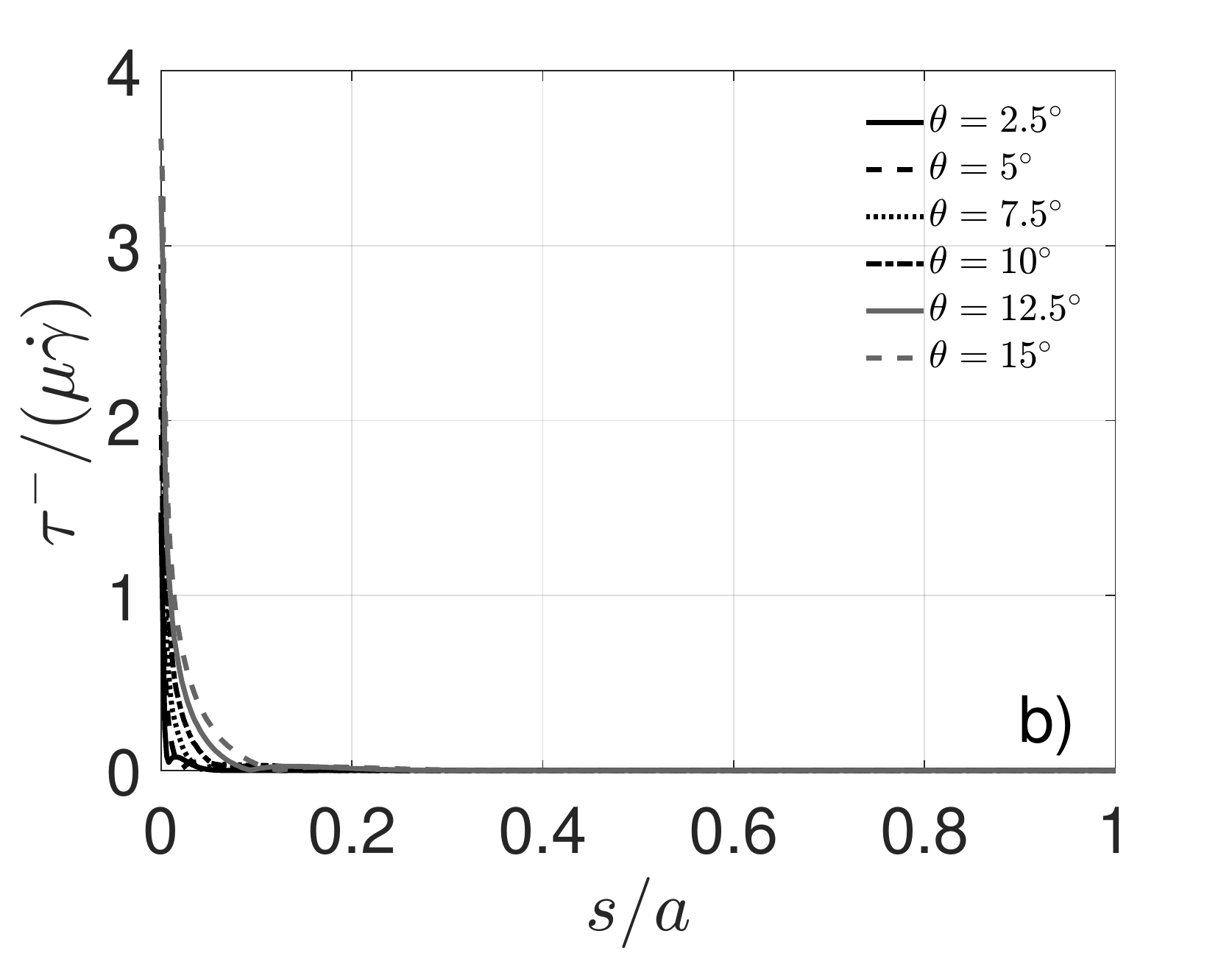}
\includegraphics[width=\textwidth]{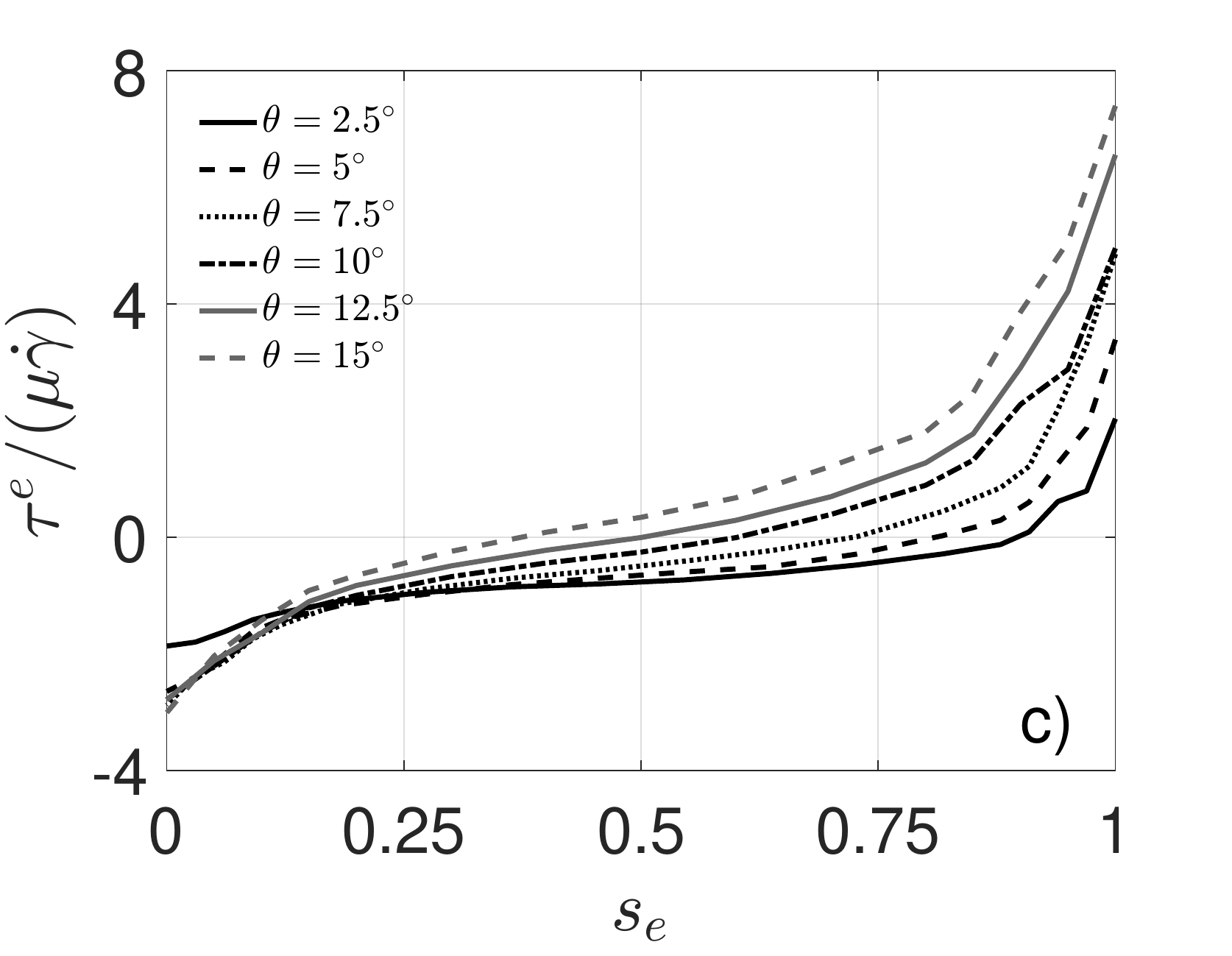}
\caption{(a) and (b): Shear stress distribution on the surfaces $\Omega^+$ and $\Omega^-$, respectively. The thickness of each layer is $h= a/50$. (c) Shear stress along the edge surface $\Omega^e$.}
\label{fgr:shearstressdistribution}
\end{minipage}
\end{figure}
 
Figures \ref{fgr:pressuredistribution} and \ref{fgr:shearstressdistribution}   show the pressure and shear stress distributions along the flap for different values of $\theta$.  In addition to providing the pressure and shear stress distribution on the upper and lower surface of the flap, we also provide the total pressure force per unit area $p=p^- - p^+$ and the total shear stress force per unit area $\tau=\tau^+ + \tau^-$ acting on the flap (our convention is that for $\tau>0$ the tangential force is directed towards the crack tip). The superscripts ``+'' and ``-'' refer to the surfaces $\Omega^+$ and $\Omega^-$, respectively. 

The  distribution of pressure and shear stress can be  separated in a near-edge region  where the hydrodynamic stresses have a large variation over a region of small $O(h)$ spatial extent, and a region far from the edge where the pressure and shear stress  vary weakly with $s$. The flow velocities are small in the wedge region so the pressure is practically constant and the shear rate is negligible. 

Because of the linearity of the Stokes flow, the pressure in the wedge region is proportional to $\mu \dot \gamma$, with a  constant of proportionality that increases with $\theta$. The signs of $p^-$ and $p^+$ in the far-edge region are such that the pressure acts to open the wedge ($p>0$). However, for sufficiently small angles, the pressure $p$ near the edge becomes negative. The shear stress on the edge $\tau_e$ acts mostly downward for small angles (see velocity field in Fig. \ref{fgr:velocity}a), pushing the flap toward the substrate. Therefore, contrary to intuition, for small angles ($\theta < \theta_c \simeq 7.5 ^\circ$) both the pressure and the shear stress at the edge act in the direction of closing the wedge. For angles larger than a critical angle $\theta_c \simeq 7.5 ^\circ$, the hydrodynamic stresses lead to wedge opening.

\begin{figure}[!htb]
\centering
\begin{subfigure}{0.9\textwidth}
\centering
\includegraphics[width=\textwidth]{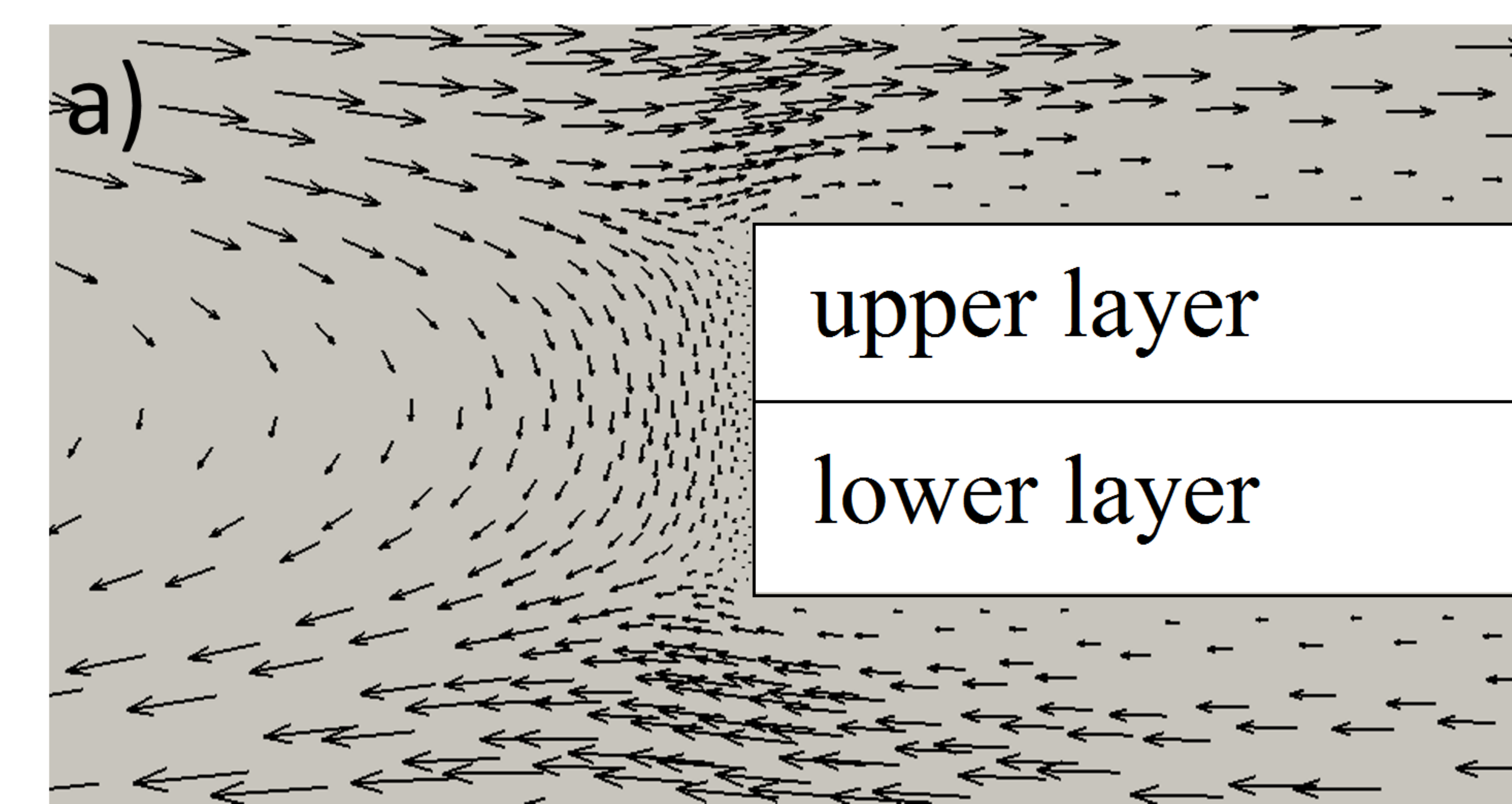}
\label{fgr:velh5}
\end{subfigure}
\begin{subfigure}{0.9\textwidth}
\centering
\includegraphics[width=\textwidth]{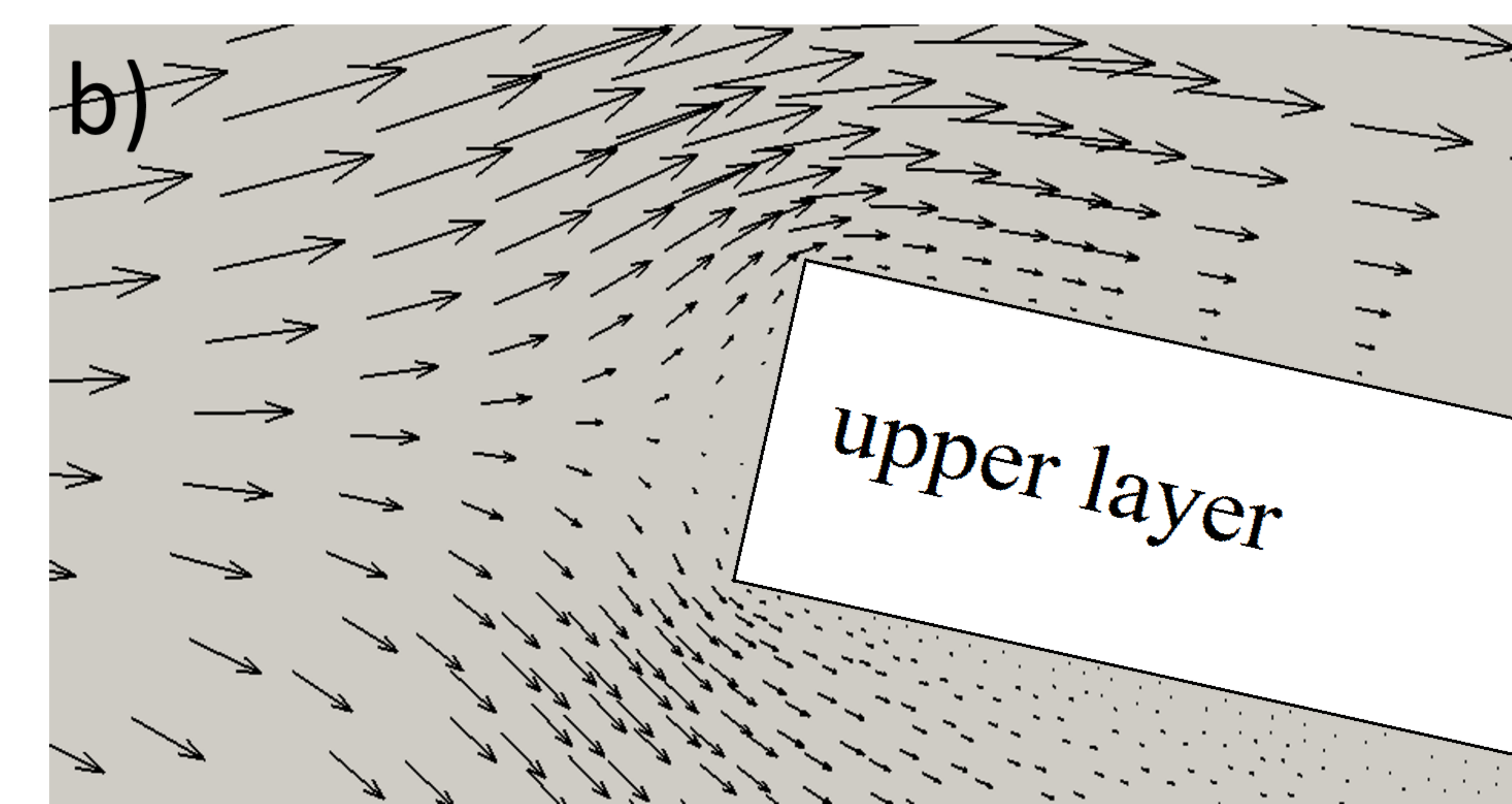}
\label{fgr:velh0_1}
\end{subfigure}
\caption{Velocity field near the edge for (a) $\theta = 0^\circ$ and (b) $\theta=12.5^\circ$. }
\label{fgr:comparisonvelocityfield}
\end{figure}

\begin{figure}[!htb]
\centering
\begin{subfigure}{\textwidth}
\centering
\includegraphics[width=0.9\textwidth]{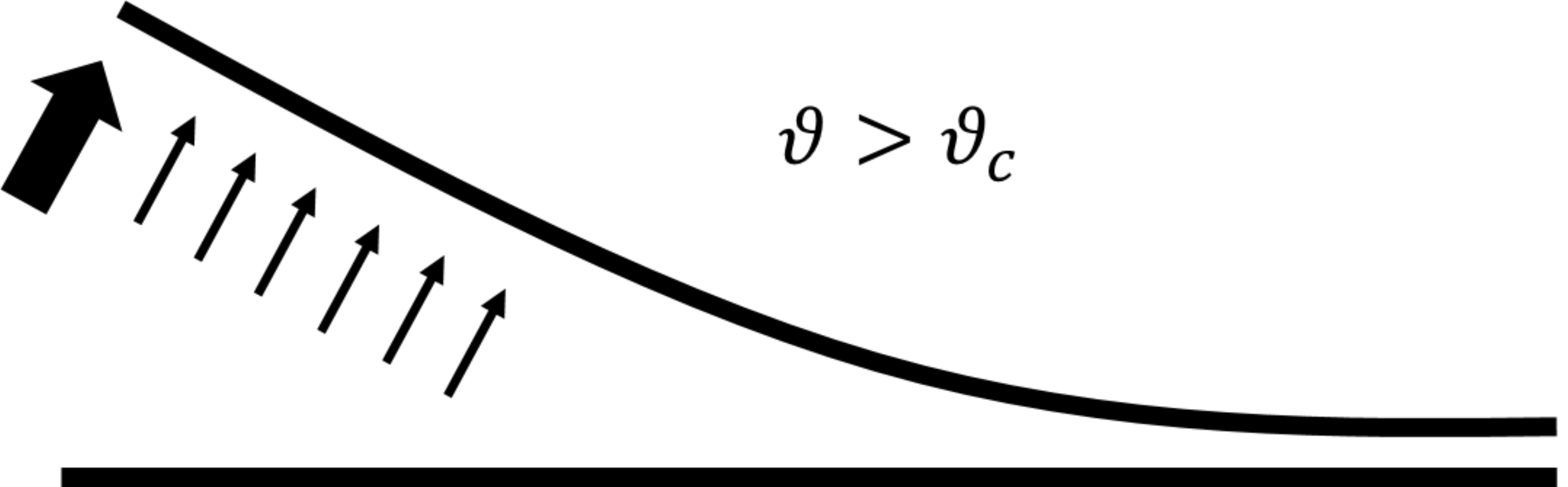}
\end{subfigure}
\begin{subfigure}{\textwidth}
\centering
\includegraphics[width=0.9\textwidth]{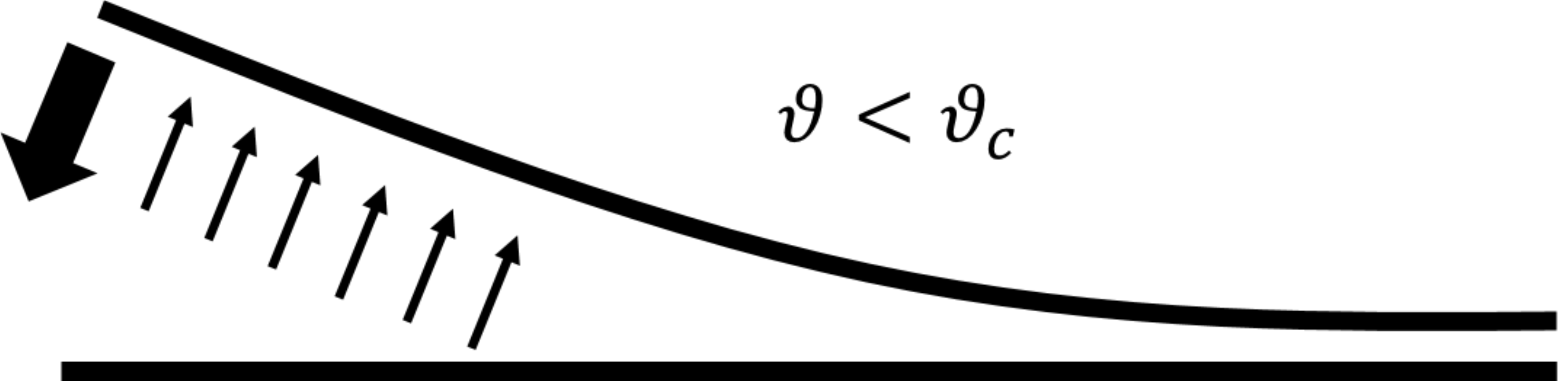}
\end{subfigure}
\caption{Schematic view of the hydrodynamic load distribution on the flap for angles larger and smaller then the critical angle $\theta_c \simeq 7.5^\circ$. }
\label{fgr:loads}
\end{figure}

To investigate why the force on the edge acts downwards for small angles, we compare in Fig.  \ref{fgr:comparisonvelocityfield} the velocity fields in the neighbourhood of the edge for $\theta = 0^\circ$ and $\theta = 12.5^\circ$. The flow field for the case $\theta = 0^\circ$ corresponds to a simulation with a horizontal flat plate of thickness $2h$, and is representative of angles much smaller than $\theta_c$. In the absence of the particle, the flow velocity would be directed from left to right  in the region $Y>0$, and from right to left in the region $Y<0$.   In the presence of the particle, for $\theta = 0^\circ$ the flow coming from the left for $Y>0$ must however change direction to satisfy the no-slip condition at the edge. This induces a flow velocity directed in the negative $Y$ direction that pushes down the flap. When $\theta > \theta_c$, the flow velocity instead points in the direction of increasing $\theta$, opening the flap (Fig. \ref{fgr:loads}). 

The existence of a critical angle is consistent with analytical results for rigid disks aligned with a shear flow \cite{singh2014rotational}. Such analysis predicts a large downward force on the edge for a thin disk immersed in a shear flow and aligned with the streamlines. It is possible that a fully two-way coupling treatment of the fluid-structure interaction problem may lead to a slightly different value of  $\theta_c$, but we believe that the existence of a critical angle is a robust result.

The implication of our results for real particles is that, in a practical setting, peeling starting from $a = 0$ would be very difficult, as the distribution of forces actually acts to close the wedge in this case. For peeling to occur, a finite edge crack of sufficient extent must exist ($a>0$), or the flap needs to present a spontaneous curvature near the edge.  In realistic cases, some of the assumptions in the model may apply only as an approximation. For example, one can expect that in instants in which the particle is inclined with respect to the flow direction a component of the hydrodynamic force would act in the direction of opening the flap. Furthermore, the edges of a real multilayer particle may in practice not be perfectly aligned. These situations require further analysis.

The shear stress assumes large positive values in correspondence to the corner points P and Q of the flap edge. The divergence of the hydrodynamic stress is a generic characteristic of flow in the vicinity of  geometrically sharp features  \citep{higdon1985stokes,pozrikidis1994shear}. Even for smoother corners, large stresses are expected near the edges with a cut-off related to the radius of curvature of the corners. In 2D nanomaterials, the curvature of the edges is cut off by a molecular scale.

The results above suggest that the essential features of the hydrodynamic load distribution are: i) an angle-dependent distributed load $Q^{\mathrm{hd}}$ on the flap, due to the effect of fluid pressure; ii) an angle-dependent edge load $F^{\mathrm{hd}}$, due to a combination of viscous shear stress and pressure. There is a further contribution due to viscous shear stress on the top surface of the flap (which also scales like $\mu \dot{\gamma}$). This stress may in principle lead to buckling, but in our situation the deformation due to the transverse load is dominant with respect to collapse due to an axial load. We will show (Figs. \ref{fgr:shape_tan} and \ref{fgr:ghat_tan}) that the inclusion of a constant tangential force 
$ \tau \simeq \mu  \dot{\gamma} $ on the flap changes the average curvature of the flap  only marginally. Hence, the inclusion of a tangential load does not change the main conclusions of our paper.

\begin{figure}[!htb]
\centering
\begin{subfigure}{\textwidth}
\centering
\includegraphics[width=0.7\textwidth]{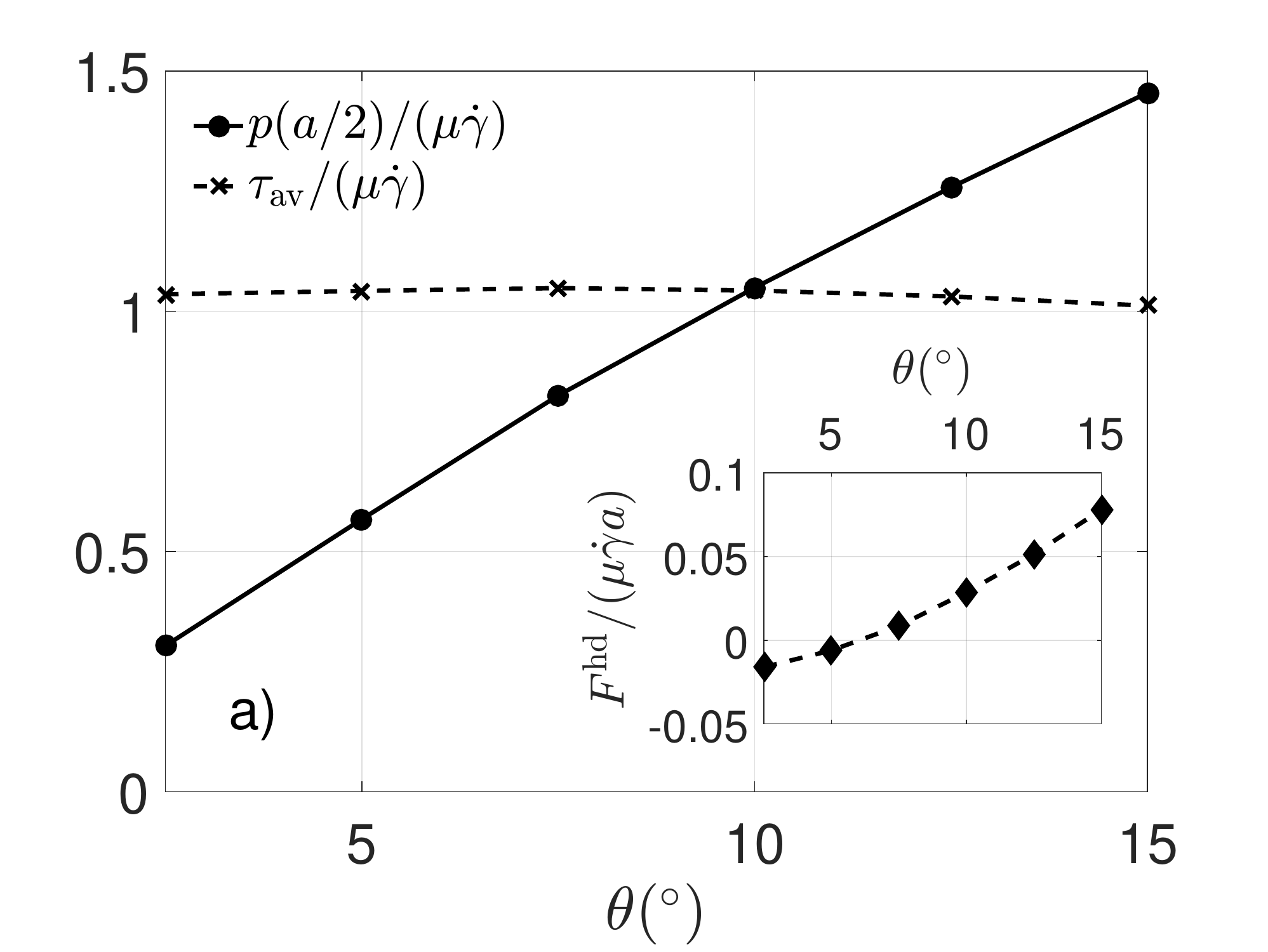}
\end{subfigure}
\begin{subfigure}{\textwidth}
\centering
\includegraphics[width=0.7\textwidth]{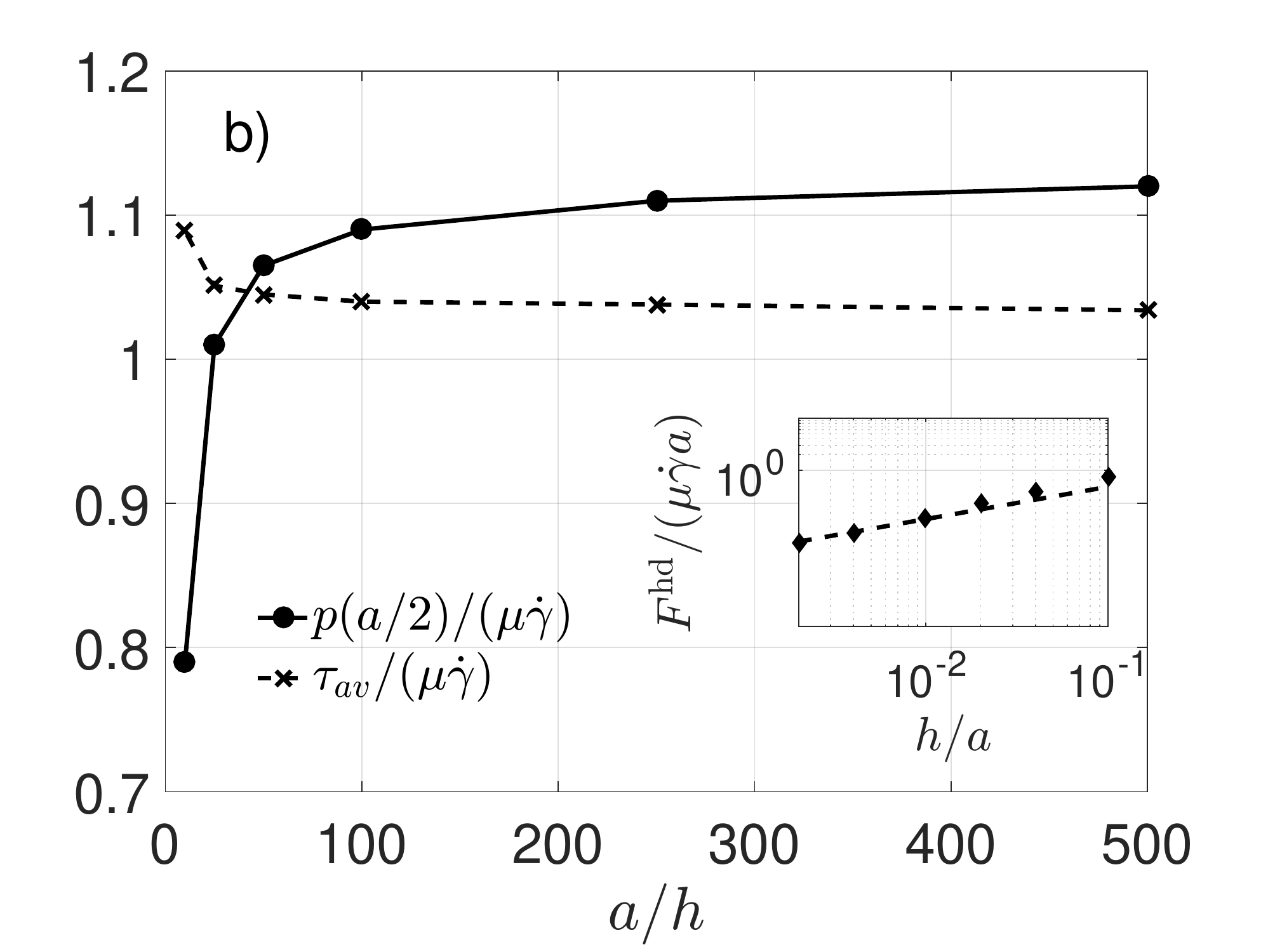}
\end{subfigure}
\caption{Total pressure force per unit area $p=p^- - p^+$ at the midpoint $s=a/2$ and  length-averaged shear stress, $\tau_{\mathrm{av}}$ as a function of (a) wedge angle for $a/h=50$ and (b) ratio $a/h$ for $\theta=10^\circ$. Inset of a): Edge force $F^{\mathrm{hd}}/(\mu \dot{\gamma}a)$ vs. $\theta$. Inset of b): Edge force $F^{\mathrm{hd}}/(\mu \dot{\gamma}a)$ vs. $h/a$ (log-log). The dashed line in the inset corresponds to a power-law exponent of $\sim 0.62$. }
\label{fig:press_shear}
\end{figure}

To quantify contribution i), we show in Fig. \ref{fig:press_shear}a  the dependence of  $p$, evaluated at the midpoint of the flap $s=a/2$, on the wedge angle $\theta$. For small angles, the linear fit 
$p(a/2)/(\mu \dot{\gamma}) = q_0+q_1\theta$, with $q_0=0.1$ and $q_1=5.37$, provides a good approximation of the simulation data. A leading order closure for the distributed hydrodynamic load is thus
\begin{equation}\label{eq:loadq}
Q^{\mathrm{hd}} = \mu \dot{\gamma} q \simeq  \mu \dot{\gamma}( q_0+q_1\theta).
\end{equation}

In contrast to the pressure load, the average shear stress (also plotted in Fig. \ref{fig:press_shear})  is almost constant when plotted against $\theta$. In Fig. \ref{fig:press_shear}b the pressure and shear stress are plotted against the thickness. As $h/a\rightarrow 0$, $p(s=a/2)$ and $\tau_{\mathrm{av}}$ become independent of the thickness.  {In the calculations presented in the current paper, we have chosen the stresses for  $a/h=50$ as representative of the thin flap limit.}

To quantify contribution ii), we show in the inset of Fig. \ref{fig:press_shear}a the dependence on $\theta$ of $F^{\mathrm{hd}}=\int_{\Omega_F}(-p + \bm{\tau} \cdot \mathbf{n}) \mathbf{n}d \ell$, where $\tau$ is the viscous stress tensor and $\Omega_{F}$ includes the surface $\Omega^e$ and  the portions of the surfaces $\Omega^+$ and $\Omega^-$ within a distance $h$ from the corner points $P$ and $Q$.  Because in Stokes flow both $p$ and $\bm{\tau}$ are proportional to $\mu \dot{\gamma}$, we can also write $F^{\mathrm{hd}} = \mu \dot{\gamma} a \tilde{F}$, where $\tilde{F}$ is independent of the fluid viscosity and shear rate.
The dependence of $\tilde{F}$ on $\theta$ shows more marked deviations from linearity than in the case of contribution i). However for small $\theta$ a linear fit, 
\begin{equation}\label{eq:loadf}
F^{\mathrm{hd}} = \mu \dot{\gamma} a \tilde{F} \simeq \mu \dot{\gamma}a(f_0 + f_1\theta), 
\end{equation}
with $f_0=-0.03$ and $f_1=0.43$  is a reasonable approximation. Equations (\ref{eq:loadq}) and (\ref{eq:loadf}) provide a linear model for the hydrodynamic load acting on the flap as a function of the configuration parameters $\theta$ and $a$, the fluid viscosity $\mu$ and the shear rate $\dot{\gamma}$. Following Ref. \citep{pozrikidis2011}, in our analysis we have neglected the effect of the hydrodynamic moment on the edge, as this contribution is negligible for very thin structures.
{On the flap edge the hydrodynamic stress at a sharp corner diverges, but the singularity is integrable \citep{michael1977separation}. As shown in the inset of Fig. \ref{fig:press_shear}b, in our finite-mesh calculations the edge force goes to zero as $h/a \rightarrow 0$ with an effective power-law exponent close to $0.62$ for small values of $h/a$. This exponent is consistent with the range of near-corner power-law stress singularity exponents reported in the literature \citep{mustakis1998microhydrodynamics}. In the current paper, we choose a reference value of $a/h=50$ to illustrate the effect of a finite edge force on the shape of the sheet, as we are interest in plausible, non-zero values of $h/a$.}

In the fluid mechanics simulation the flap is straight. Therefore, for a given value of $a$, the configuration is parametrised by a unique value of $\theta$. But how do we relate the opening angle in the solid mechanics calculation to the one in the fluid mechanics calculation?  The angle $\theta$ has to be approximated as a function of the flap shape. Among the possible approximations, one could use the angle at the tip of the flap, $\theta \simeq\de w_1/\de x|_{x=0}$, the local angle $\theta \simeq \de w_1/\de x$, or the secant angle made by the secant line (connecting the flap tip to the crack tip) with the horizontal,  $\theta \simeq  w_1(0)/a$. For $\theta \ll 1$,  $ w_1(0)/a\simeq  \de w_1/\de x|_{x=0}$, and the difference between using the local angle or the secant angle is small. We choose the secant angle approximation $\theta = w_1(0)/a $ in the small displacement model, since it is typically used when the opening angle varies slowly \citep{snoeijer2006free} and gives particularly simple analytical solutions. The effect of using different approximations for $\theta$ will be analysed in the context of the large displacement model.

\subsection{Solid mechanics model}
\label{sec:analytical}
\begin{figure}[htbp]
\centering
\includegraphics[width=0.9\textwidth]{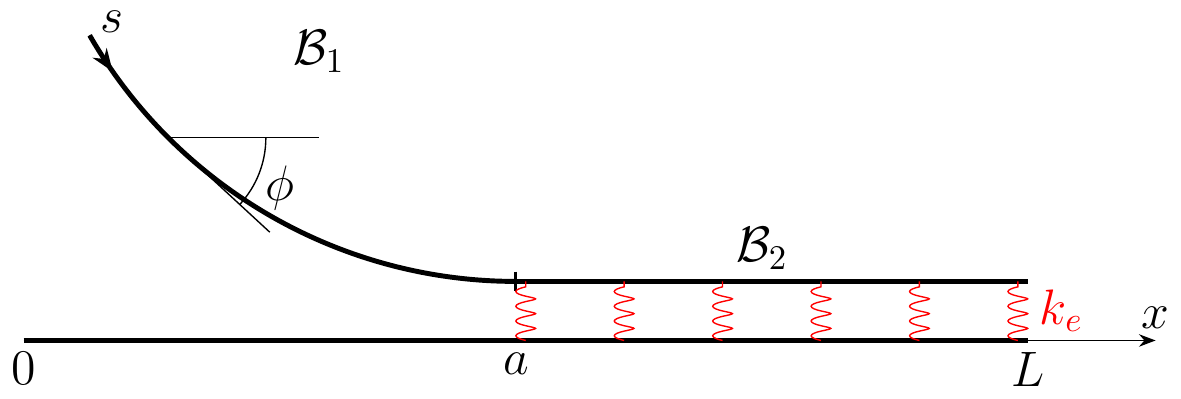}
\caption{Schematic of the solid mechanics model. }
\label{fgr:mechmod}
\end{figure}
The solid mechanics model uses the closures for the loads $F^{\mathrm{hd}}$  and $Q^{\mathrm{hd}}$ obtained in the previous section to calculate the elastic deformation of the flap.  In the model, we neglect tangential loads. We will show later that tangential loads make only a negligible contribution to the deformation of the flap.  We consider both a small displacement model (which we solve analytically) and a large displacement model. As shown in Fig. \ref{fgr:mechmod}, the deformable layer is divided in two regions: the region $\mathcal{B}_1$ on which the hydrodynamic load is applied, and the bonded region $\mathcal{B}_2$ in which the hydrodynamic load is zero. The out-of-plane  displacements corresponding to   $\mathcal{B}_1$ and  $\mathcal{B}_2$ are indicated by $w_1(x)$ and $w_2(x)$, respectively. 

We initially consider both a small-displacement model, valid for $|\nabla w_1| ~\ll~ 1$. Later, we compare against a large-displacement model. In the small displacement model, the inter-layer interface  is modelled  as an elastic foundation \emph{à la} Winkler \citep{wang2005beams}, characterized by a foundation modulus $k_e$. For $|\nabla w_1| ~\ll~ 1$, $w_1$ satisfies  
\begin{equation}
D\, \dfrac{\de^4 w_1}{\de x^4} = Q^{\mathrm{hd}} 
\label{eq:B1}
\end{equation}
where $x$ ranges from the coordinate corresponding to edge of the flap ($x=0$) to the crack tip ($x=a$), and $D$ is the bending stiffness. The  equation for $w_2$ is 
\begin{equation}\label{eq:B2}
D\, \dfrac{\de^4 w_2}{\de x^4} + k_e\, w_2 = 0.   
\end{equation}
The boundary condition at $x=0$ requires $D\de^3 w_1/\de x^3|_{x=0} ~=~  F^{\mathrm{hd}}$ and $ D\de^2 w_1/\de x^2|_{x=0}~=~ M^{\mathrm{hd}}$ where $F^{\mathrm{hd}}$ and $M^{\mathrm{hd}}$ are the hydrodynamic force and moment acting on the edge, respectively. Assuming that $L\gg a$, the boundary conditions at infinity satisfy  $w_2(x\rightarrow\infty)=0$ and $\de w_2/\de x|_{x\rightarrow\infty}=0$.  The solutions for $\mathcal{B}_1$ and  $\mathcal{B}_2$ are matched by enforcing continuity of the out-of-plane displacement and its derivatives at $x=a$ up to the third order (see Eqs. (\ref{eq:cont1})-(\ref{eq:cont4}) in \ref{sec:appendixA}). 

The consideration of a soft foundation in the small-displacement model adds to the generality of the results. Furthermore, the interlayer interface in 2D nanomaterials does not necessarily correspond to an infinitely stiff foundation, because the range of the interlayer force and the size of the cohesive zone is nanometric but so can be relevant displacements. For instance, the analysis of the case with the elastic foundation could be useful to interpret molecular dynamics results, where the range of maximum flap deflection and crack length (a few nanometres) is not necessarily orders of magnitude larger than the size of the cohesive zone ( $\left(Dd_0^2/\Gamma\right)^{1/4} \sim 1\,\mathrm{nm}$, using typical parameters for single-layer graphene). Molecular dynamics results of peeling in liquids are now appearing which could benefit from our analysis \citep{sresht2015,bordes2017,bordes2019}. We are carrying out similar molecular dynamic investigations in our group as well.

In the large-displacement model, we solve a non-linear equation for the curvature of the region $\mathcal{B}_1$ of the deformable layer.  The equations of equilibrium of forces and moments for an inextensible elastica with a purely normal follower load $Q^{\mathrm{hd}}$ are 
\begin{equation}
\dfrac{\de^2 M}{\de s^2}-\kappa N-Q^{\mathrm{hd}}=0 \label{eq:force_mom1}
\end{equation}
and
\begin{equation}
\kappa\dfrac{\de M}{\de s}+\dfrac{\de N}{\de s}=0 \label{eq:force_mom2},
\end{equation}
respectively \citep{antman1968general}. Here, $s$ is the curvilinear coordinate along the flap,  $\phi$ is the tangent angle to the flap,  $\kappa=-\de \phi/\de s$ is the curvature,  $M=D \kappa$ is the bending moment and $N$ is the axial (internal) force.  Integration of Eq. (\ref{eq:force_mom2}) gives $D\kappa^2/2 + N = c$, where $c$ is a constant. Evaluating this constant at $s=0$ (the flap edge) gives $ N(s) = N^\mathrm{hd}+ (M^\mathrm{hd})^2/(2D)  - D \kappa^2/2$, where we have used the boundary conditions $M^\mathrm{hd} = D \kappa(0)$ and $N^\mathrm{hd}$ is an axial force applied to the free end. Substituting into Eq. (\ref{eq:force_mom1}) yields
\begin{equation}
D\dfrac{\de ^2\kappa}{\de s^2} - \left (N^\mathrm{hd} + \frac{\left (M^\mathrm{hd} \right)^2}{2D} \right) \kappa  +\frac{D}{2}\kappa^3-Q^{\mathrm{hd}}=0.
\end{equation} 
In the analysis for large displacements we neglected the effect the axial load $N^\mathrm{hd}$ and the hydrodynamic moment on the edge $M^\mathrm{hd}$. The equation governing the flap shape reduces to   
\begin{equation}
D\dfrac{\de ^2\kappa}{\de s^2}+\frac{D}{2}\kappa^3-Q^{\mathrm{hd}}=0.
\label{eq:B1_largedisplacements}
\end{equation} 
To limit the number of cases, in the large-displacement  analysis, we did not include the Winkler's foundation and assumed that the flap is clamped at $s=a$, corresponding to the boundary condition $\phi(a) = 0$. We also neglected the normal load $F^\mathrm{hd}$ applied on the edge, imposing instead free end boundary conditions $\kappa(0) = 0$ and $\de\kappa/\de s|_{s=0} = 0$.   The shape of the flap was calculated from $x=a +\int_a^s \cos\phi(s)\de s=a-\int_0^a \cos\phi(s)\de s+\int_0^s \cos\phi(s)\de s$ and $y=\int_a^s\sin\phi(s)\de s=-\int_0^a\sin\phi(s)\de s+\int_0^s\sin\phi(s)\de s$.

\subsubsection{Analysis of flap shape and critical shear rate}  
\label{sec:res_solid}

In the small-displacement analysis, we derive analytical solutions to (\ref{eq:B1}) and (\ref{eq:B2}), and compare against numerical solutions. The numerical solutions were obtained with a finite difference scheme, approximating the derivatives at interior points using second-order, central differences  and using skew operators at the boundaries \citep{tornberg2004simulating}; the resulting discrete system was solved  by matrix inversion. In the large displacement analysis, we only discuss finite-difference solutions of Eq. (\ref{eq:B1_largedisplacements}), seen as an equation for $\phi$. The non-linear system was solved by a Newton-Raphson method. 

The critical fluid shear rate to initiate fracture  of the  inter-layer interface  is calculated using Griffith's energy balance, assuming brittle fracture. Denoting by $\Gamma$ the total solid-solid adhesion energy per unit area  (i.e. twice the solid-solid surface energy), the condition for crack initiation according to Griffith's theory is 
\begin{equation}\label{eq:Griffith}
G = \Gamma
\end{equation}
where  $G = \dfrac{\partial U}{\partial a}$ is the strain energy release rate (\citep{griffith1921}) and $U$ is the bending energy  per unit length:
\begin{equation}
U=\frac{D}{2}\int_0^L  \kappa^2 \de s \simeq \frac{D}{2}\int_0^a  \left(\dfrac{\de^2 w_1 }{\de x^2}\right)^2 \de s+\frac{D}{2}\int_a^L  \left(\dfrac{\de^2 w_2 }{\de x^2}\right)^2 \de s.
\end{equation}

Recasting the equilibrium equation for the flap and Griffith's balance  into non-dimensional variables, using $a$ and $D$ to scale the other variables (see \ref{sec:appendixA} for the small displacement formulation), makes it evident that the initiation of the crack is controlled by  three non-dimensional parameters:
\begin{equation}
\ghat= \frac{\mu\, \dot{\gamma}\, a^3}{D} \qquad \widehat {\Gamma}=\frac{\Gamma a^2}{D} \qquad  \chi^4= \frac{k_e\, a^4}{4\,D} 
\label{eq:par}
\end{equation}
The first parameter, the \textit{non-dimensional shear rate}, is the ratio of hydrodynamic forces and bending forces. The second parameter, the \textit{non-dimensional adhesion energy}, is the ratio of adhesion and bending forces.  The parameter $\chi$ represents the ratio between the crack length $a$ and the cohesion length $\lambda = (4D/k_e)^{1/4}$. An infinitely stiff interlayer interface corresponds to $\chi\rightarrow\infty$. For a brittle-like law $k_e = 2\,\Gamma /d_0^2$, $\chi$ can be rewritten as $\chi^4=\Ghat/2({a/d_0})\,^2$, where $d_0$ is a molecular scale characterising the range of the adhesion forces ($d_0 \simeq 1\,\mathrm{nm}$). 

In our analysis we consider relatively small wedge angles. In the fluid mechanics simulations we consider  at most a $\theta \approx 15^\circ$. In the solid mechanics simulations we extrapolate the results to larger angles, but still assuming that $\theta$ is significantly smaller than $\pi/2$. Based on our numerical experiments, this condition on the angle roughly corresponds to  $\mu\, \dot{\gamma}\, a^3/D<1$. For these values of the non-dimensional shear rate the flap does not buckle, and maintains a qualitative shape similar to that in Fig. \ref{fgr:flap}. 

Typical values for the surface energy $\Gamma/2$ of graphene in vacuum or inert gases are around $0.1\,\mathrm{N/m}$ ($0.115\, \mathrm{N/m}$ \citep{van2017}, $0.085\, \mathrm{N/m}$ \citep{chen2012vortex}, $0.070\, \mathrm{N/m}$ \citep{hernandez2008high}, $0.047\, \mathrm{N/m}$\citep{wang2009wettability}). In a very controlled adhesion experiment using a modified force balance apparatus, Engers et al. (\citep{van2017}) recently reported, in the case of single-layer graphene, $0.115 \pm 0.004\,\mathrm{N/m}$ for dry nitrogen, $0.083 \pm 0.007\,\mathrm{N/m}$ for water, and $0.029 \pm 0.006\,\mathrm{N/m}$ for sodium cholate, a surfactant recommended for liquid-phase exfoliation processes. N-methylpyrrolidone (NMP) is considered an optimal solvent for graphene exfoliation. Molecular dynamics studies \citep{shih2010} suggest that NMP reduces the specific interaction energy between graphene nanosheets as compared to water by a factor of about 2 (from $\approx 250\,\mathrm{kJmol^{-1} nm^{-2}}$ for water to $\approx 110-120\,\mathrm{kJ mol^{-1} nm^{-2}}$ for NMP). Although more research is needed to clarify the effect of solvent on adhesion during crack initiation in 2D nanomaterials, it seems from the data above that good solvents can reduce the adhesion energy significantly, but this reduction is probably not by several orders of magnitude.  Values between $\Gamma = 0.1 \,\mathrm{N/m}$ and  $\Gamma = 0.01 \,\mathrm{N/m}$ are probably realistic.   

We discuss the small-displacement results for two  cases : 
\begin{itemize}
\item Case 1: distributed load only ($Q^{\mathrm{hd}} \neq 0; F^{\mathrm{hd}}=0$); 
\item Case 2: distributed load plus edge load ($Q^{\mathrm{hd}} \neq 0; F^{\mathrm{hd}}\neq 0$). 
\end{itemize}
The analytical derivations are conceptually simple, but rather cumbersome. The quadratic dependence of the bending energy on the displacement gives rise to many coupling terms, and going through the derivation step by step may obscure their physical meaning. Here we report the main results, particularly focusing on the structure of the solution.  The complete derivations are reported  in \ref{sec:appendixA} and \ref{sec:appendixA1}.

\textit{Case 1, angle-independent load, infinitely stiff foundation}. The solution is the classical solution for a cantilever beam subject to a constant load: 
\begin{equation}\label{eq:w1_cant}
w_1(x)=\frac{\ghat q_0}{24 a^3}(x-a)^2(x^2+2ax+3a^2).
\end{equation} 
The corresponding non-dimensional bending energy is 
\begin{equation}\label{eq:U_cant}
\widehat{U}=\frac{q_0^2\ghat\,^2}{40},
\end{equation} 
and the critical shear rate (from Eq. (\ref{eq:Griffith})) is 
\begin{equation}\label{eq:G_cant}
\ghat_c=\frac{2\sqrt{2}}{q_0}\Ghat^{\frac{1}{2}}.
\end{equation}
Because the load is constant, the bending energy is quadratic in $\ghat$. As a consequence, the non-dimensional critical shear rate depends on the square root of the non-dimensional adhesion parameter. 

\begin{figure}[htb]
\centering
\includegraphics[width=0.9\columnwidth]{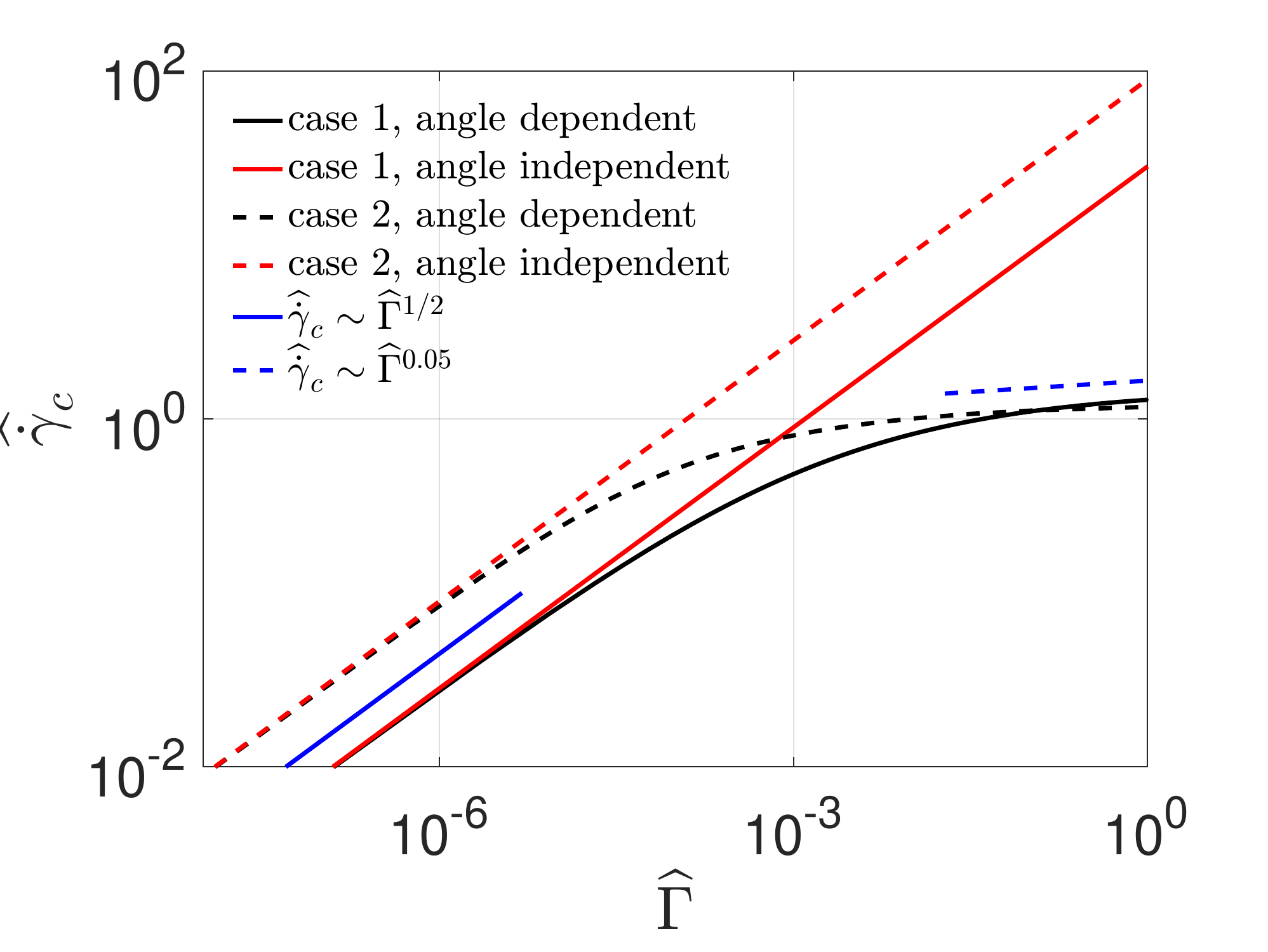}
\caption{Non-dimensional shear rate as a function of the non-dimensional adhesion energy for different loads. The blue lines show the power-law trends for small values of $\Ghat$ (continuous line) and large values of $\Ghat$ (dashed line).}
\label{fgr:qbar}
\end{figure}

\textit{Case 2, angle-independent load, infinitely stiff foundation}.  The displacement is 
\begin{equation}\label{eq:w1_cant_f0}
w_1(x)=\frac{\ghat}{24a^3}(x-a)^2 \left (q_0(x^2+2a x+3a^2)+4af_0(x+a) \right),
\end{equation}
the dimensionless bending energy is 
\begin{equation} 
\widehat{U}=\frac{\ghat\,^2}{120}(20 f_0^2 + 15 f_0 q_0 + 3 q_0^2),
\end{equation} 
and the critical shear rate is  
\begin{equation}\label{eq:G_cant_f0}
\ghat_c=\frac{2\sqrt{2}}{q_0}\frac{1}{\sqrt{1+5f_0/q_0+20/3(f_0/q_0)^2}}\Ghat^{\frac{1}{2}}.
\end{equation}
Because we are here considering edge and distributed loads that are independent of the wedge angle, we again recover a power-law with an exponent $1/2$. The critical shear rate decreases as the hydrodynamic coefficient $q_0$ increases, by an amount that depends on the edge load coefficient $f_0$. In particular, the critical shear rate decreases as $f_0$ increases. In our case $f_0$ is negative, so the required shear rate is slightly larger than if only the distributed load was included  (see Fig. \ref{fgr:qbar}). 

\textit{Case 1 \& 2, angle-independent load, ``soft foundation''}. If $\chi$ has a finite value, the displacements in the free  and adhered portions of the flap are coupled. This brings about a dependence of the solution on $\chi$, which in turn depends on $\Ghat$ for a fixed $d_0/a$. The critical shear rate in case 1 is 
\begin{equation}\label{eq:G_soft}
\ghat_c=\frac{2\sqrt{2}}{q_0}\Ghat^{\frac{1}{2}}\left(\frac{\chi}{1+\chi}\right)^{3/2}.
\end{equation}
A similar expression holds for case 2, with a numerical prefactor  now depending on $f_0$ (see \ref{sec:appendixA1}, Eq. (\ref{eq:app_Gsoft_f0})).

\begin{figure}[htbp]
\centering
\includegraphics[width=0.8\textwidth]{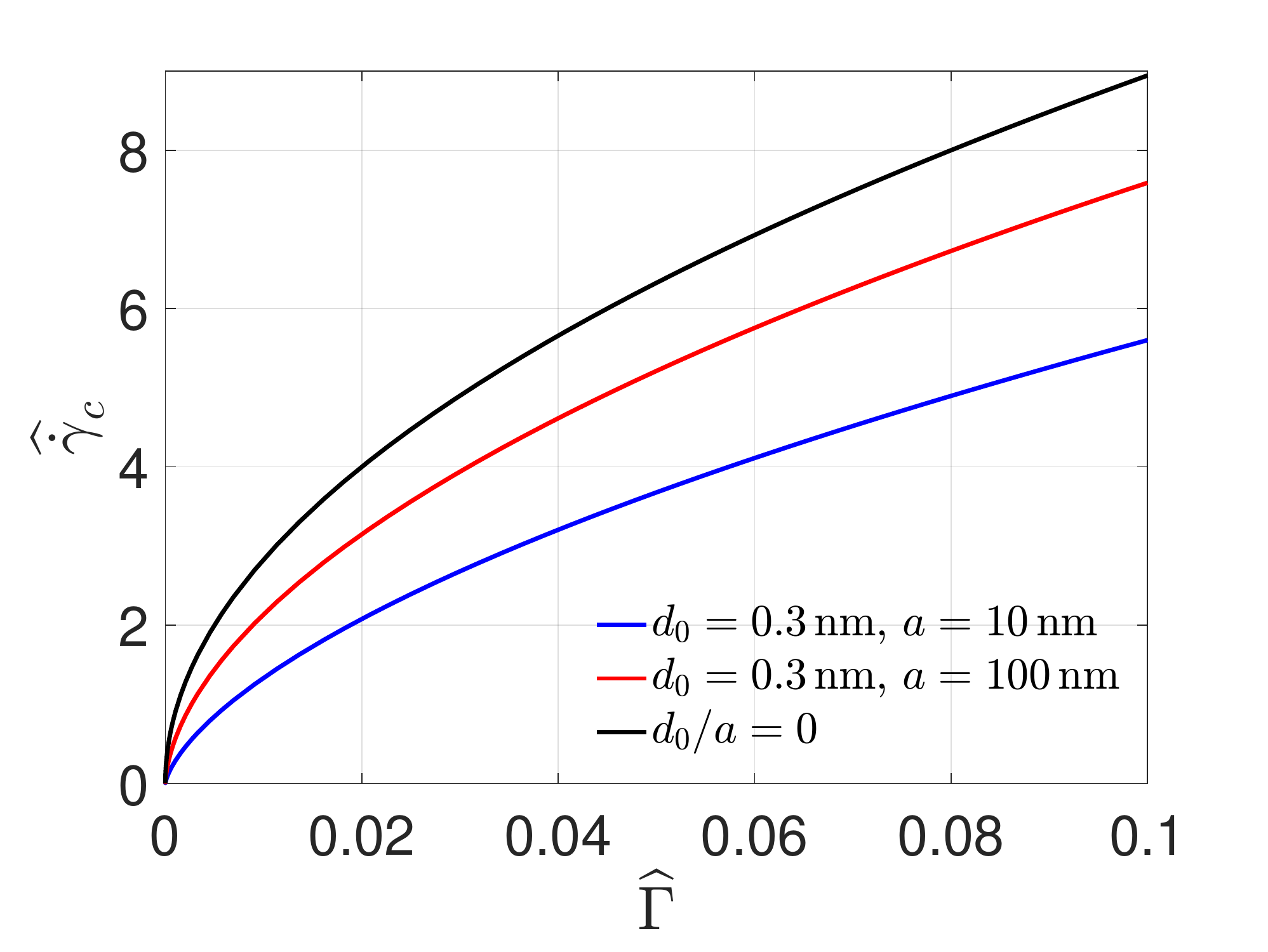}
\caption{Non-dimensional shear rate $\ghat_c$ as a function of the non-dimensional adhesion energy $\Ghat$ for different values of the parameter $d_0/a$. An increase in the stiffness of the foundation (smaller $d_0$) corresponds to larger values of the critical shear rate}
\label{fgr:chi_finite}
\end{figure}
While the load is constant, owing to the coupling of the flap deformation to the mechanics in the adhered portion of the flap, the relation between shear rate and adhesion energy is not a power law. We typically expect $\chi \gg 1$, so deviations from a power law behaviour are small. By plotting the critical shear rate in log-log scale,  the data can be fitted to an effective power-law exponent, whose value depends on the specific value of $d_0/a$.  Fig. \ref{fgr:chi_finite} shows $\ghat_c$ as a function of $\Ghat$ for different values of $a$ and $d_0=0.3\,\mathrm{nm}$. Since the exponent for soft foundations is larger than for rigid foundations, the critical shear rate decreases as the foundation becomes less stiff. From Eq. (\ref{eq:G_soft}) we can see that $\ghat_c\propto\Ghat^{7/8}$ for $\chi\rightarrow 0$ and $\ghat_c\propto\Ghat^{1/2}$ and $\chi\rightarrow \infty$.  The effective power-law exponent is therefore bounded between $1/2$ and $7/8$, with higher shear rates corresponding to stiffer foundations.
The boundary condition at the crack tip can be assumed to be clamped provided that $\chi \gg 1$.  For $\chi = 1$ the cohesion length $\lambda \sim (Dd_0^2/\Gamma )^{1/4}$ is of the same order of the crack length. For typical parameters, the cohesion length is of the order of $1\,\mathrm{nm}$ for single-layer graphene, and up to a few nanometres for few-layer graphene. The soft foundation case examined here can therefore be useful to interpret molecular dynamics results, where due to computational constraints the crack length is typically at most $10-20\,\mathrm{nm}$ \citep{bordes2019}.

\textit{Case 1 \& 2, angle dependent load, infinitely stiff foundation}. The consideration of a dependence on $\theta$ now introduces a non-linear dependence of $w_1$ on $\ghat$. This dependence is particularly simple to analyse when $\theta$ is approximated as the secant angle.  In this case, the flap displacement and bending energy expressions, for case 1, are given by  
\begin{equation}\label{eq:w1_q0q1}
w_1(x)=\frac{\ghat q_0 }{3a^3 (8 - \ghat q_1)}(x-a)^2(x^2+2ax+3a^2). 
\end{equation}
and
\begin{equation}\label{eq:U_q}
\widehat{U}=\frac{8q_0^2\ghat\,^2}{5(8-q_1\ghat)^2},
\end{equation} 
respectively. The requirement $w_1 \geq 0$ means that these equations are valid for $\ghat\leq 8/q_1 \simeq 1.49$; the requirement of a positive solution is consistent with our initial assumption $\ghat < 1$. 

\begin{figure}[htb]
\centering
\includegraphics[width=0.8\textwidth]{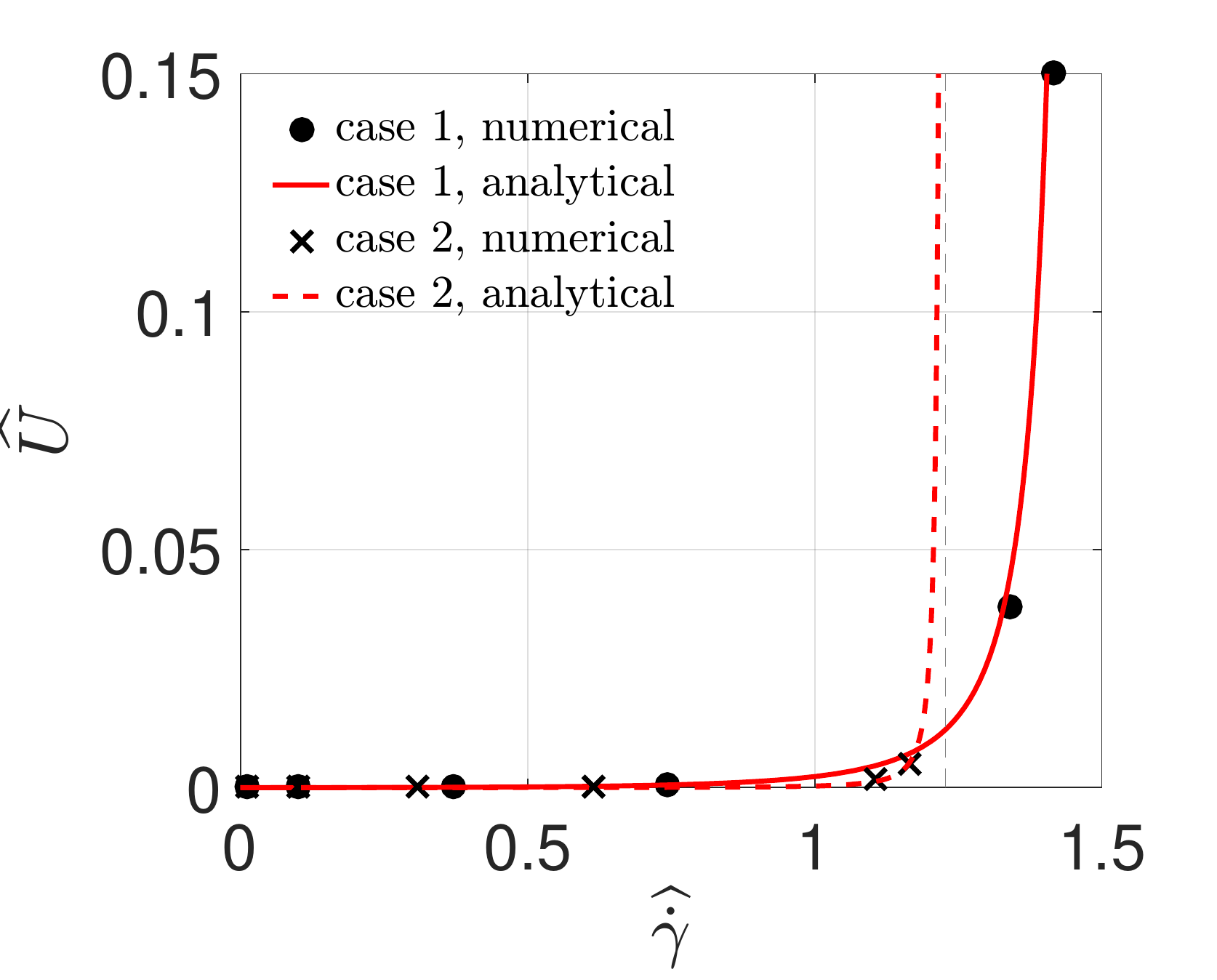}
\caption{Non-dimensional bending energy as a function of the non dimensional shear rate. The markers represent the results from finite difference simulations. The red lines represent the analytical solutions.}
\label{fgr:U}
\end{figure}

There is an interesting difference with respect to the angle-independent case. Expression (\ref{eq:w1_q0q1}) displays the same dependence on the variable $x$ as the corresponding solution for an angle independent load, Eq. (\ref{eq:w1_cant}). However the prefactor diverges as $\ghat$ approaches a finite value $8/q_1$. The corresponding bending energy expression,
displays, expectedly, the same divergence. As we will see in the analysis of the large displacement case, this divergence is a robust feature (although different approximations to $\theta$ give somewhat different  values of $\ghat$ for which the flap  curvature diverges). This divergence is important as it will completely change the dependence of the critical shear rate on the non-dimensional adhesion energy. Case 2 also displays a divergence at a slightly different value of the shear rate.  The presence of an edge load  gives 
\begin{equation}\label{eq:w1_qf}
w_1(x)=\frac{\ghat(x-a)^2(\ghat\left(f_1q_0-f_0q_1\right)(2x^2+ax)-6q_0(x^2+2ax+3a^2)-24af_0(x+2a))}{6a^3(\ghat(8f_1+3q_1)-24)}.
\end{equation}  
A term ($f_1q_0-f_0q_1$) coupling the edge and  distributed load coefficients appears  at denominator, and the solution shows a divergent behaviour for $\ghat=24/(8f_1+3q_1)$. The bending energy profiles for cases 1 and 2 are plotted as a function of $\ghat$ in Fig. \ref{fgr:U}. Because $24/(8f_1+3q_1)< 8/q_1$, the presence of the edge load reduces the critical value of $\ghat$. The divergence appears slightly more sharp in case 2 than in case 1.

How is the critical shear rate related to the non-dimensional adhesion energy when the load depends on the wedge angle? In case 1, the  relationship between $\ghat$ and $\Ghat$ is 
\begin{equation}\label{eq:G_q0q1}
\frac{8}{5}q_0^2\ghat\,^2\frac{q_1\ghat+40}{(8-q_1\ghat)^3}=\Ghat.
\end{equation}
One could develop approximate solutions of this implicit equation to calculate $\ghat$ as a function of $\Ghat$, but it instead more convenient to plot $\Ghat$ as a function of $\ghat$ and then switch the axis. The result is shown in Fig. \ref{fgr:qbar}, where the angle-dependent load cases are compared to the angle-independent ones (including both cases 1 and 2). 

In the angle-dependent cases, a horizontal plateau  in the critical shear rate emerges as $\Ghat \rightarrow 1$. The plateau is particularly evident in case 2 in the range $\Ghat=10^{-3}-1$. In this range, the critical shear rate does not follow a power-law. However, if we insist on fitting a power-law to the data near $\Ghat=1$, we obtain an exponent of $0.05$, much smaller than the exponent $1/2$ obtained for $\Ghat \ll 1$. The solution thus changes behaviour, and a regime where the critical shear rate starts becoming only weakly dependent on $\Ghat$ emerges. This observation has important practical implications for the optimisation of liquid-exfoliation processes, as discussed in the conclusions section.    

\begin{figure}[htb]
\centering
\includegraphics[width=0.9\columnwidth]{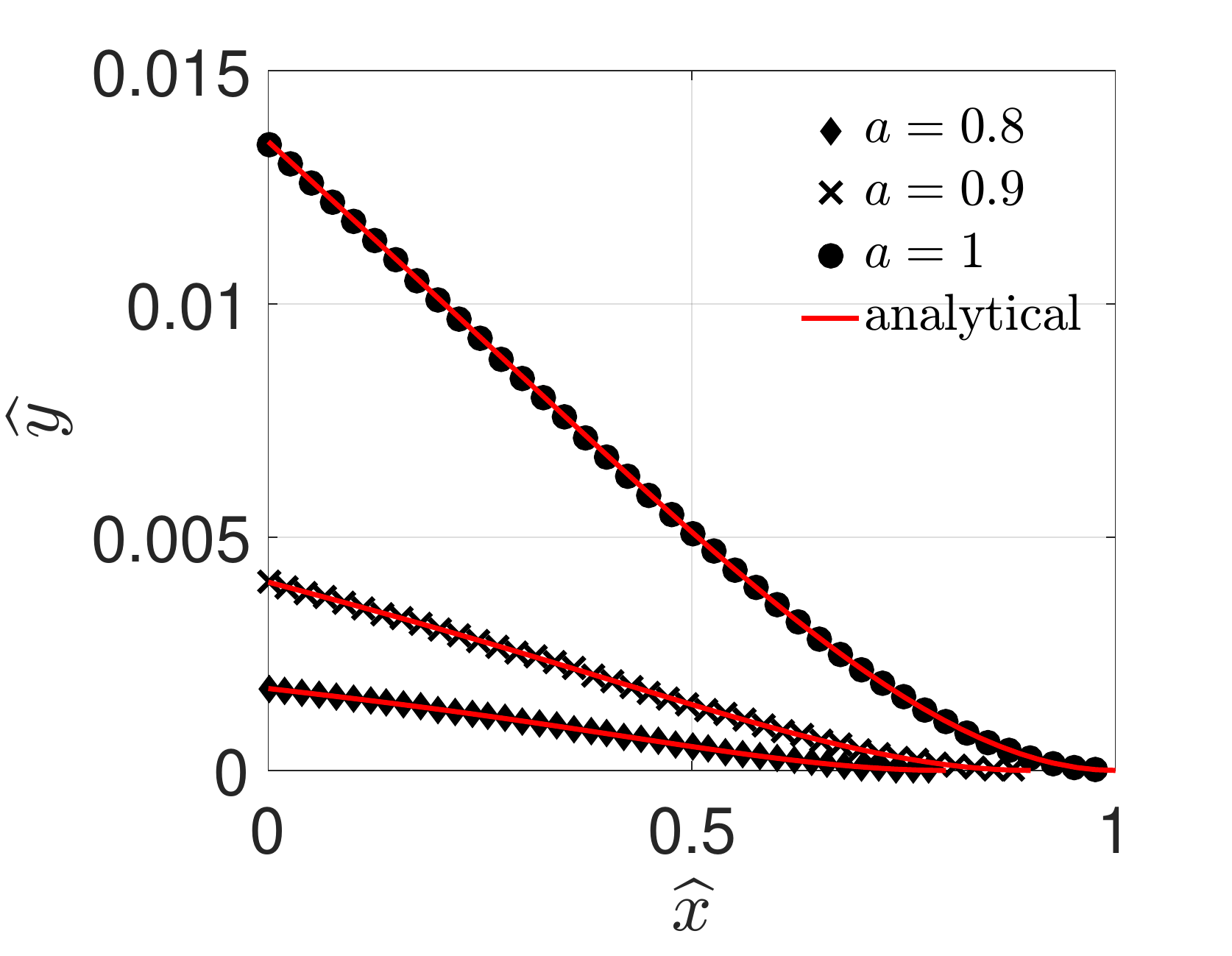}
\caption{Out-of-plane displacement of the flap from simulations (markers) and analytical solutions (red lines) for three values of $a$, angle dependent load in case 2 (edge load plus distributed load) and $\mu\dot\gamma/D=1$. The results are in units of $(D/(\mu\dot\gamma))^{1/3}$ . }
\label{fgr:edgeload}
\end{figure}

Figure \ref{fgr:edgeload} illustrates the deformation of the flap for three different values of $a$ (i.e. three different values of $\ghat$). Both the edge and the distributed loads are considered, as well as the dependence on $\theta$. The red lines indicate analytical solutions, while the markers indicate  numerical results. For small values of $a$ (the smallest value of $a$ considered is $a=0.8$), the edge load is negative and bends the tip of the flap slightly downwards. When $a$ increases (or, equivalently, $\ghat$ increases) this effect becomes less evident as the distributed load becomes dominant. 

The action of the edge load opposing the opening of the wedge determines a smaller deformation of the flap if compared with the deformation without edge load. For relatively small values of $\Ghat$,  the larger curvature of the flap in case 1 causes $\ghat$ to be smaller than in case 2 (compare continuous black line and dashed black line in Fig. \ref{fgr:qbar}). This difference decreases as $\Ghat$ increases and $\ghat$ approaches the asymptotic value $8/q_1-24/(8f_1+3q_1)$. We therefore conclude that the edge load is quantitatively relevant for small values of $\Ghat$ or, equivalently, of $a$ (i.e. at the initial stages of the peeling). The inclusion of the edge load in the model requires higher values of $\ghat$ to sustain the peeling mechanism and avoid the closure of the wedge. For larger values of the $\Ghat$ or $a$, i.e. larger deformations, the edge load can be neglected.

\textit{Large displacement model}. We now discuss numerical predictions based on the large-displacement model. Given that the divergence in the bending energy that gives rise to the plateau seen in Fig. \ref{fgr:qbar} is due to large curvatures, it is natural to enquire whether the results hold if non-linear terms in the equation governing the flap shape are retained. We focus on the case that includes only the distributed load, as we have shown that the effect of the edge load  is  important only for small values of $a$. 

\begin{figure}[htb]
\centering
\includegraphics[width=0.9\columnwidth]{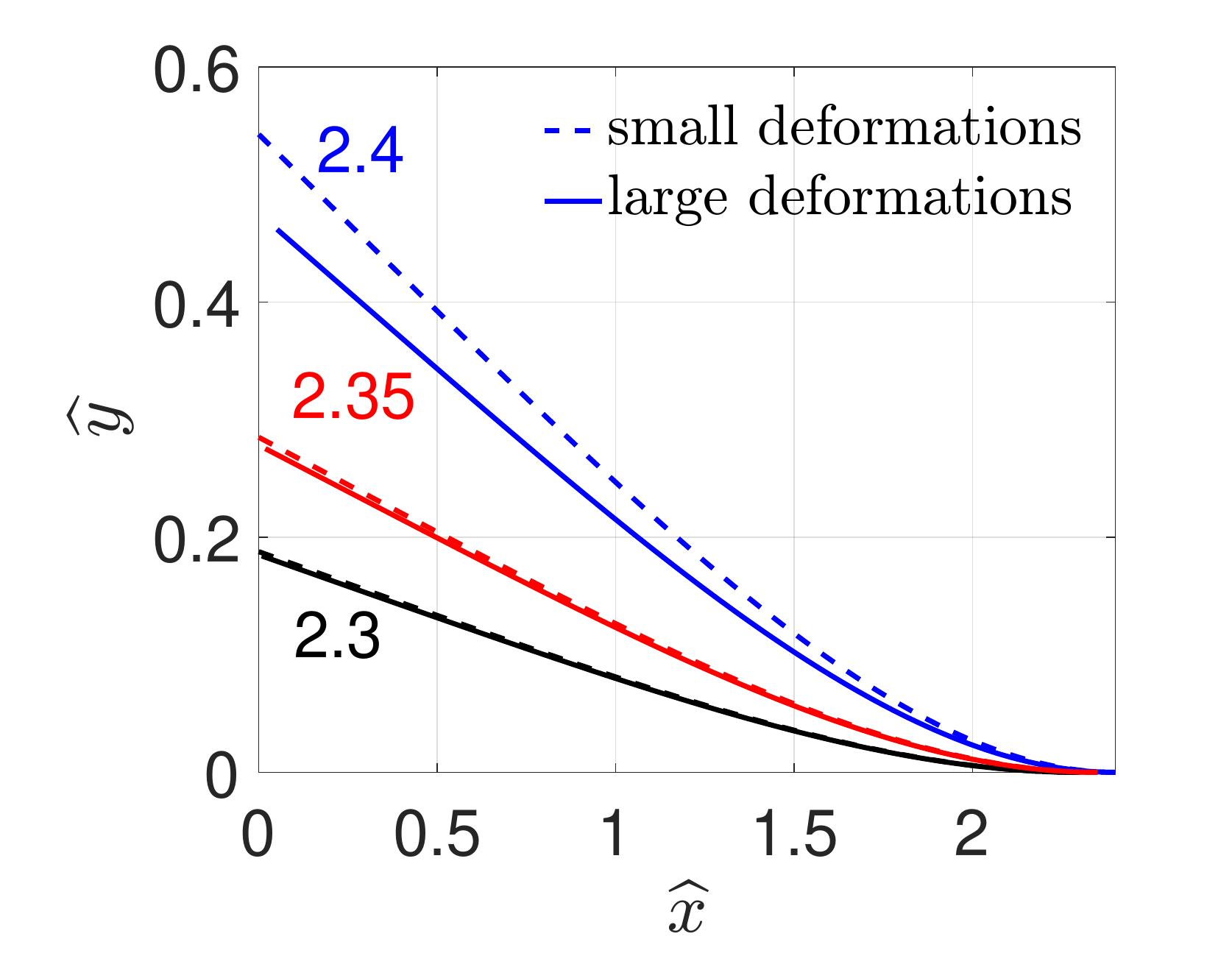}
\caption{Flap shape comparing the linear (small displacement) and the non-linear (large displacement) formulation. The flap is plotted for different values of $a$ and $\mu\dot\gamma/D=0.1$. The quantities in the plot are in $(D/(\mu\dot\gamma))^{1/3}$ units).}
\label{fgr:peel0}
\end{figure}

Figure \ref{fgr:peel0} compares numerical results for the flap shape obtained using Eq. (\ref{eq:B1_largedisplacements}) with those obtained with Eq. (\ref{eq:B1}).  Appreciable deviations due to non-linearity occur for $a\simeq 2.30$ (in units of $(D/(\mu\dot\gamma))^{1/3}$ ), corresponding to $\ghat \simeq 1.217$. This value is quite close to the value $\ghat = 8/q_1 \simeq 1.49$ for which the bending energy diverges in the linear formulation.  The largest deviations are more evident near the edge of the flap.  However, the high curvature in the region near the crack tip is well captured by the small displacement theory even for $a=2.4$, corresponding to $\ghat \simeq 1.382$. 

\begin{figure}[htbp]
\centering
\includegraphics[width=0.9\textwidth]{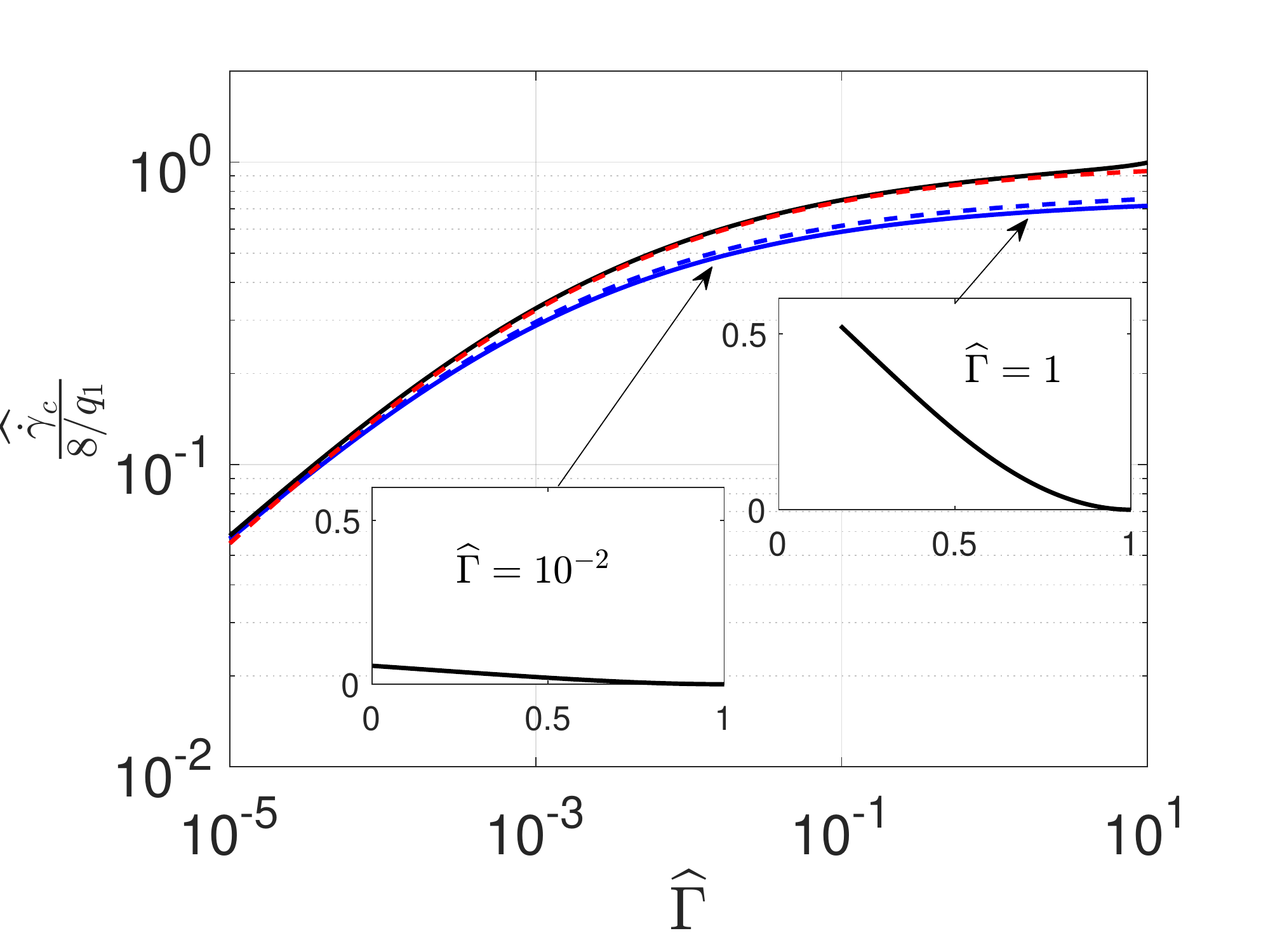}
\caption{ Non-dimensional shear rate as a function of the non-dimensional adhesion energy in the large displacement theory for different approximations of the wedge angle: tip angle (continuous blue line), local angle (dashed blue line) and secant angle (black line). The red line is the analytical solution for small displacements and secant angle approximation. }
\label{fgr:shearrate12}
\end{figure}

Because the flap is practically straight far from the crack tip, the value of the bending energy is dominated by the curvature near the crack tip, for which the linear formulation appears to give reasonably accurate results. As a consequence we expect the critical shear rate predicted by the linear and non-linear theories to display comparable trends. 

In Fig. \ref{fgr:shearrate12} the critical shear rate  is plotted in log-log scale against the non-dimensional adhesion energy. In addition to comparing linear and non-linear deformation theories, we also show results for different approximation of the wedge angle. The non-linear theory using the secant angle follows closely the corresponding linear one, giving only slightly larger values. For example, for $\Ghat=2$ the value of $\ghat_c$ given by the non-linear theory (in the secant angle approximation) is only about $2\%$ larger than the corresponding value in the linear theory.  

\begin{figure}[htbp]
\centering
\includegraphics[width=0.9\textwidth]{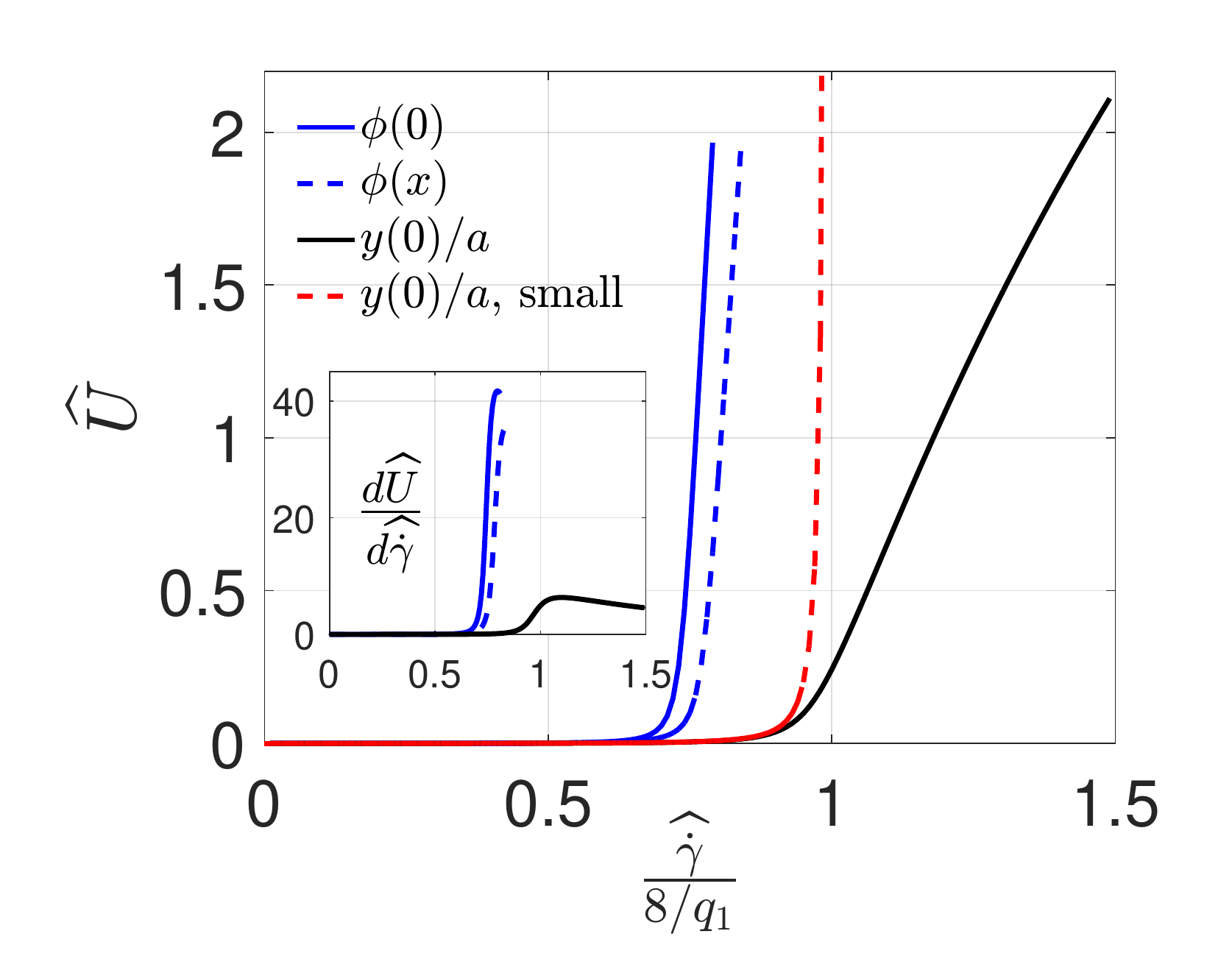}
\caption{Non-dimensional bending energy vs. non-dimensional shear rate in the large displacement theory for different approximations of the wedge angle in the forcing term. The red dashed line correspond to the small-displacement theory, while the other lines are for the large-displacement theory. The inset shows the derivative of $\widehat{U}$ with respect to $\ghat$. }
\label{fgr:energy12}
\end{figure}

 As shown in Fig. \ref{fgr:energy12}, different approximations to the wedge angle essentially change the value for which the bending energy diverges. Correspondingly, the curves $\Ghat - \ghat$ are shifted upwards or downwards depending on the specific approximation for the wedge angle adopted (recall that a vertical asymptote in the $\hat{U}-\ghat$ curve corresponds to a horizontal plateau in the $\ghat-\Ghat$ curve).  From Fig. \ref{fgr:energy12}, we can see that the effect of including non-linear terms is essentially to make the divergence less sharp. This result is confirmed by the inset in Fig. \ref{fgr:energy12} showing the non-divergence of the  derivative of $\widehat{U}$. 
 
 We could not derive explicit analytical expressions for the full non-linear equation. A linear equation that captures large displacements more accurately than Eq. (\ref{eq:B1}) is obtained from Eq. (\ref{eq:B1_largedisplacements}) by setting the term proportional to $\kappa^3$ to zero. Using the definition of the curvature $\kappa=-\de\phi/\de s$ and the local angle approximation $Q^{\mathrm{hd}}=\mu\dot\gamma (q_0+q_1\phi)$, we obtain a linear equation in the rotation:
\begin{equation}\label{eq:phi}
D\dfrac{\de^3 \phi}{\de s^3}+\mu\dot\gamma(q_0+q_1\phi)=0
\end{equation}
The solution is 
\begin{equation} \label{eqn:sol}
\phi(s)=\frac{q_0}{q_1}\left(-1+\frac{e^{(a-s)/l}+2e^{(2a+s)/(2l)}\cos\left(\frac{\sqrt{3}s}{2l}\right)}{1+2e^{3a/(2l)}\cos\left(\frac{\sqrt{3}a}{2l}\right)}\right)
\end{equation}
where $l=D/(\mu\dot\gamma q_1)^{1/3}$ is the length scale of the exponential decay of the curvature from the crack tip. Equation (\ref{eqn:sol}) shows a divergence when the denominator approaches zero (i.e. for $\ghat\cong 1.18$ when $a=1$). Comparing this analytical solution against the solution of the full non-linear equation  shows that the divergence is only slightly mitigated by the term depending on the cube of the curvature. 

\begin{figure}[htb]
\centering
\includegraphics[width=0.9\columnwidth]{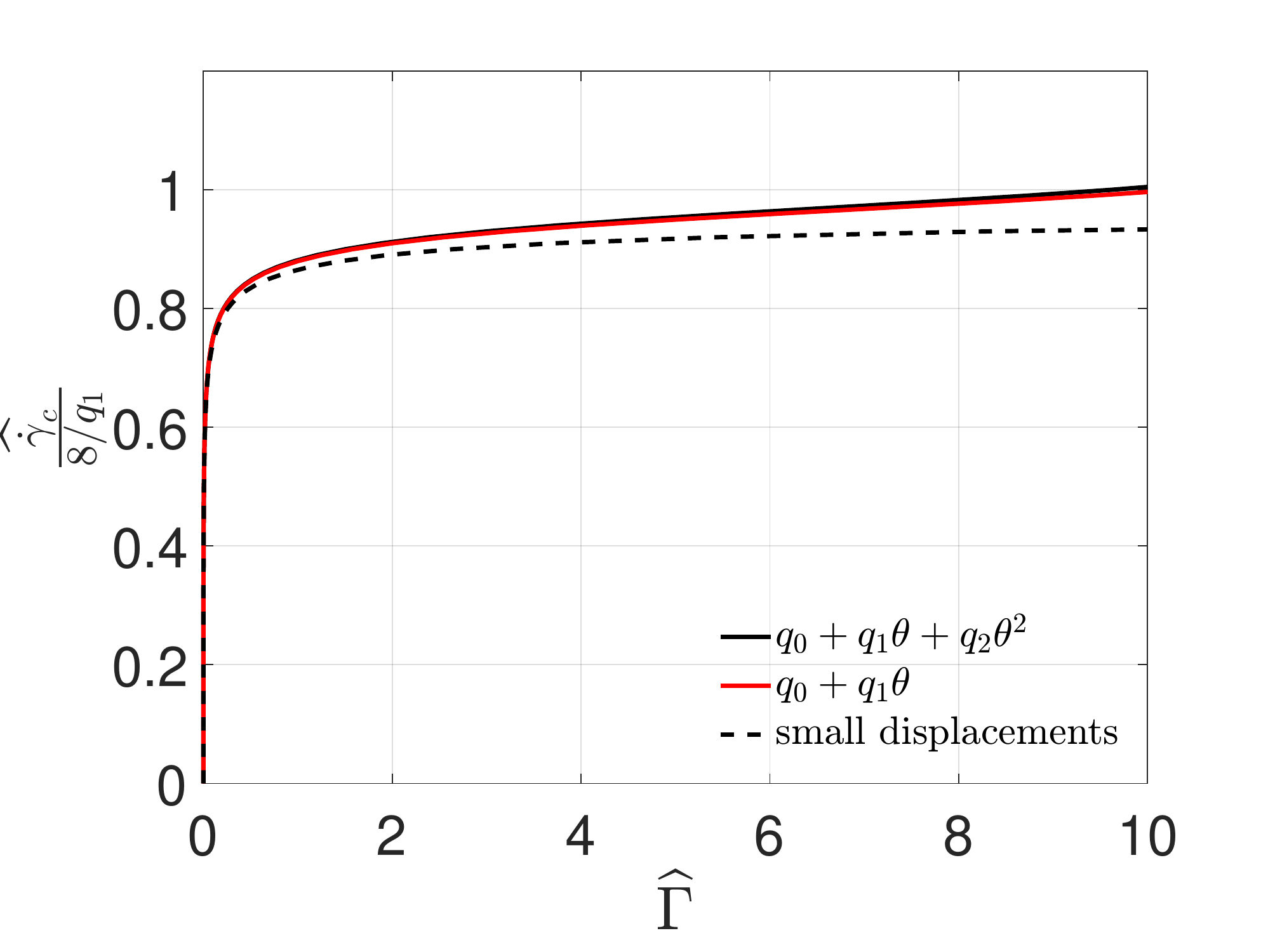}
\caption{Relation between non-dimensional shear rate and non-dimensional adhesion energy, comparing the linear and  non-linear loads.}
\label{fgr:nonlinearload}
\end{figure}

\textit{Effect of non-linear load and tangential stress}. In our solutions, we have considered a load that depends linearly on the wedge angle. A closer observation of Fig. \ref{fig:press_shear}a shows a slight downward curvature in the plot of the distributed load. In our range of parameters, considering non-linear variations of the form $q=q_0 + q_1 \theta + q_2 \theta^2$, where $q_2<0$ (a best fit to the flow simulation data gives $q_0=0.1$, $q_1=5.37$ and $q_2=-0.07$),  changes the behaviour of the solution only very marginally. Because the quadratic term is negative, the load rises less than linearly with the angle. As a consequence, the critical shear rate is slightly higher than if the quadratic term is neglected (Fig. \ref{fgr:nonlinearload}). Nevertheless, this downward curvature is an interesting feature, because we expect that for large angles at some point the hydrodynamic load will decrease. The small effects that we see in the current section will therefore be amplified, potentially changing the behaviour of the solution.

\begin{figure}[htb]
\centering
\includegraphics[width=0.9\columnwidth]{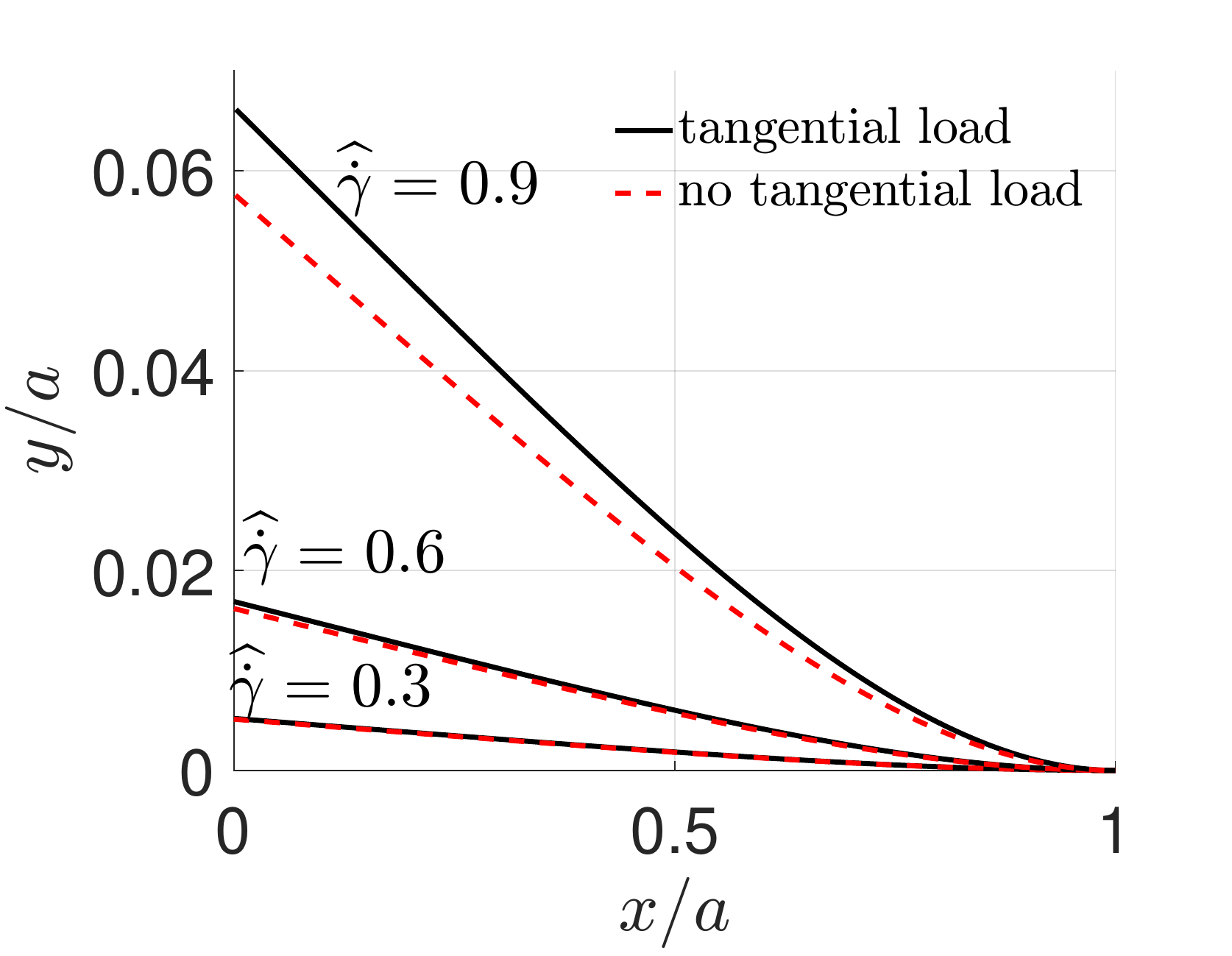}
\caption{Flap shape in the large displacement limit with tangential load (black line) and without tangential load $\tau =1.05 \mu\dot\gamma $ (red dashed line) for different values of $\ghat$.}
\label{fgr:shape_tan}
\end{figure}

\begin{figure}[htb]
\centering
\includegraphics[width=0.9\columnwidth]{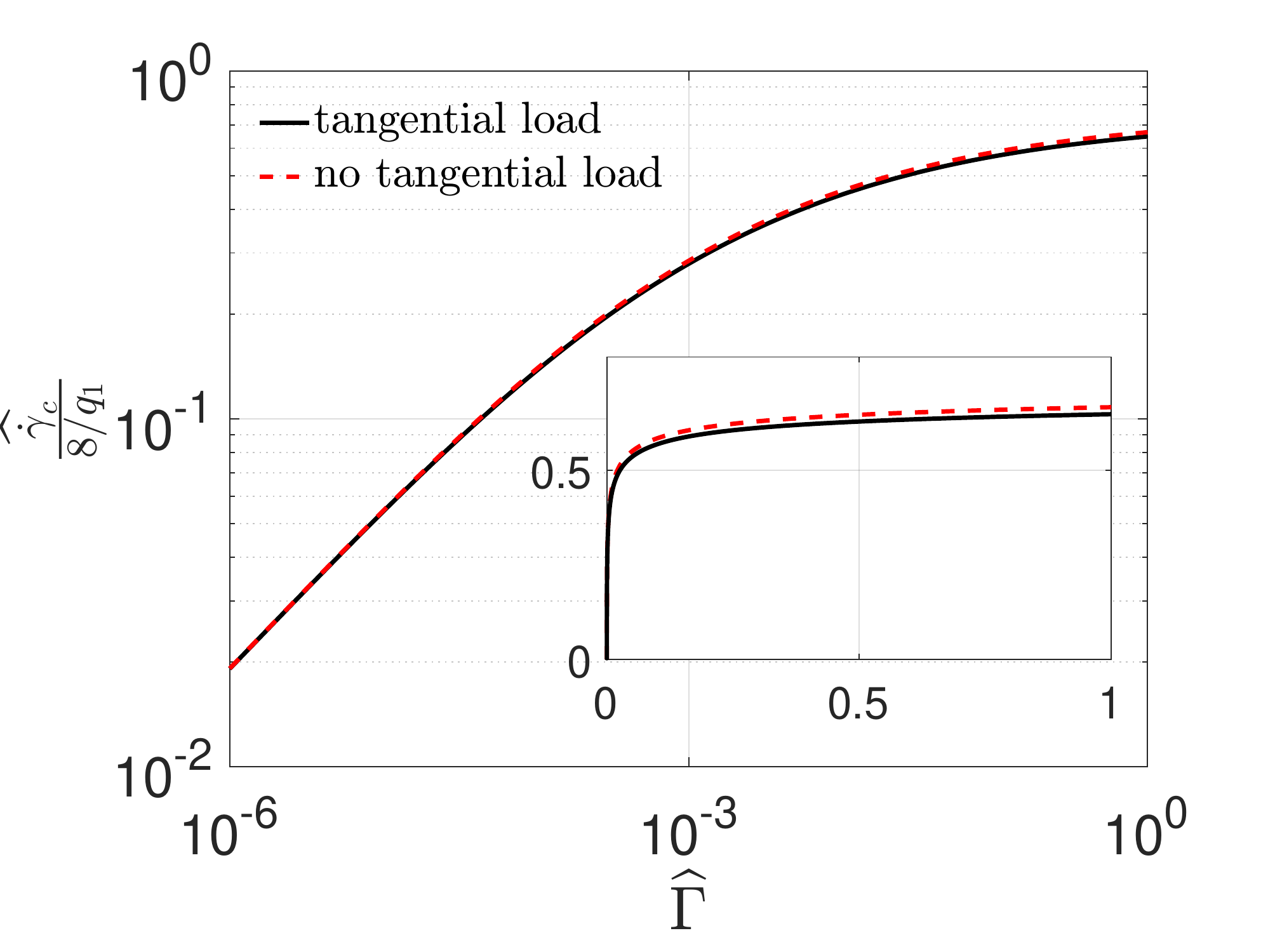}
\caption{Non-dimensional shear rate against non-dimensional adhesion energy in the large displacement limit with tangential load $\tau =1.05 \mu\dot\gamma $ (black line) and without tangential load (red dashed line) applied on the flap. The inset shows the same plot on a linear scale.}
\label{fgr:ghat_tan}
\end{figure}

In our analysis, we have also neglected the tangential distributed load, although this is of order $\sim \mu \dot{\gamma}$ as the normal load, under the assumption that bending of the flap originates mostly from normal loads for relatively stiff flaps.  We have found that, if this assumption is removed by accounting for a uniform tangential load in the large displacement model  (this was done by modifying Eq. (\ref{eq:force_mom2}) to account for a constant $\tau \simeq\mu\dot\gamma $), the flap shape is altered but not to an extent as to change the main conclusions drawn so far. We show in Fig. \ref{fgr:shape_tan} the shape of the flap for two simulations, with and without the tangential load, and for different values of  $\ghat$. As  $\ghat$ increases, the effect of including the tangential load on the maximum displacement becomes more marked. However, the curvature near the tip seems to be largely independent of the presence of the tangential load. As a consequence, the critical shear rate when the tangential load is accounted for is only slightly smaller than when only normal loads are used (Fig. \ref{fgr:ghat_tan}). In the small displacement model, the axial deformation does not influence the curvature, hence the tangential distributed load does not influence the energy balance and the critical shear rate. In the large displacement model, the normal and tangential components are coupled but the tangential load does not affect the curvature drastically, as we have just seen.

Within the assumptions of our model a straight solution is not an equilibrium solution, because a finite pressure also acts for $\theta=0$ (extrapolated result, see Fig. \ref{fig:press_shear}a). Even for nearly straight flaps, the deformation is due mostly to  the transverse load.  Axial and transverse loads in our problem are not independent:  increasing the shear stress on the top surface of the flap also causes an increase in the pressure below the flap. Thus, the transverse deformation due to transverse load occurs before a classical buckling instability sets in.

\textit{Regimes of exfoliation}. Our analysis suggests that the dependence of the load on the flap configuration, a purely hydrodynamic effect, gives rise to a transition in the relation between the non-dimensional critical shear rate and the non-dimensional adhesion parameter. When $\Ghat$ is truly infinitesimal, $\Ghat \ll 10^{-5}-10^{-4}$, the dependence of the load on the configuration is small ($q_1 \theta \ll q_0$) and $\ghat_c\sim\Ghat^{1/2}$.  However, for larger values of  $\Ghat$ the opening angle increases (inset of Fig. \ref{fgr:shearrate12}), and the dependence of the load on $\theta$  becomes important ($q_1 \theta \sim q_0$). In this regime, the flap displacement is not proportional to the shear rate, and a plateau emerges in which $\ghat_c$ is at most a weak function of $\Ghat$. The transition occurs for quite small opening angles. Setting $q_1 \theta = q_0$, we get $\theta \simeq 0.0186$, which corresponds to about $1^\circ$.

In dimensional terms, the order of magnitude of the critical shear rate in the two regimes is 
\begin{equation}
\dot\gamma_c\sim \frac{(\Gamma D)^{1/2}}{\mu a^2}
\label{eqn:regime1}
\end{equation}
and
\begin{equation}
\dot{\gamma_c} \sim\frac{D}{\mu a^3} f,
\label{eqn:regime2}
\end{equation}
respectively, where $f=O(1)$ is at most a weak function of $\Ghat$. Mathematically, the weak dependence on adhesion in the intermediate range of values of $\Ghat$ can be understood by looking at Eq. (\ref{eq:G_q0q1}). We rewrite this equation  as $8q_0^2\ghat\,^2(q_1\ghat/5+8)=\Ghat[8-q_1\ghat]^3$. An increase in $\Ghat$ would give an increase in $\ghat$ if the term in square parenthesis was neglected. But  an increase in $\ghat$ corresponds  also to a decrease in the factor $(8-q_1\ghat)$ in the right hand side of the equation. The two terms on the right-hand side therefore compensate each other, leading to the asymptotic behaviour that is only weakly dependent on $\Ghat$.

The critical shear rate is predicted to depend on the initial size of the crack, but not on the particle size $L$ directly. This results follows from the assumption that the cohesion zone is smaller than the length of the solid-solid interface, an assumption that is expected to hold in practice. If the crack size correlates with the particle size (for example, if $a = c_1 L$ in a statistical sense, with $c_1 \ll 1$), then Eqs. (\ref{eqn:regime1}) and (\ref{eqn:regime2}) should be used with $L$ replacing $a$ and changing the prefactor accordingly.

Our conclusions are valid up to $\Ghat \sim 1$. For larger values of $\Ghat$,  the flap is almost vertical and our assumptions for the load fail. We expect that for  $\Ghat$ significantly larger than one the critical shear rate should start growing again.  Large values of $\Ghat$ correspond to relatively large values of $a$, so our conclusions hold for the initial development of the crack.

\section{DISCUSSION AND CONCLUSIONS}
We have proposed and analysed a model for the exfoliation of layered 2D nanomaterials suspended in a turbulent flow.  The model is based on the idea that exfoliation occurs through an erosion process, whereby layers of 2D nanomaterials are removed almost `layer-by-layer'' through a microscopic flow-induced peeling process. The model  provides insights into the dependence of the critical shear rate on the geometric, mechanical and adhesion  parameters, for a realistic hydrodynamic load distribution. For this dependence, we provide explicit analytical formulas when possible.

\begin{figure}[htb]
\centering
\includegraphics[width=0.9\columnwidth]{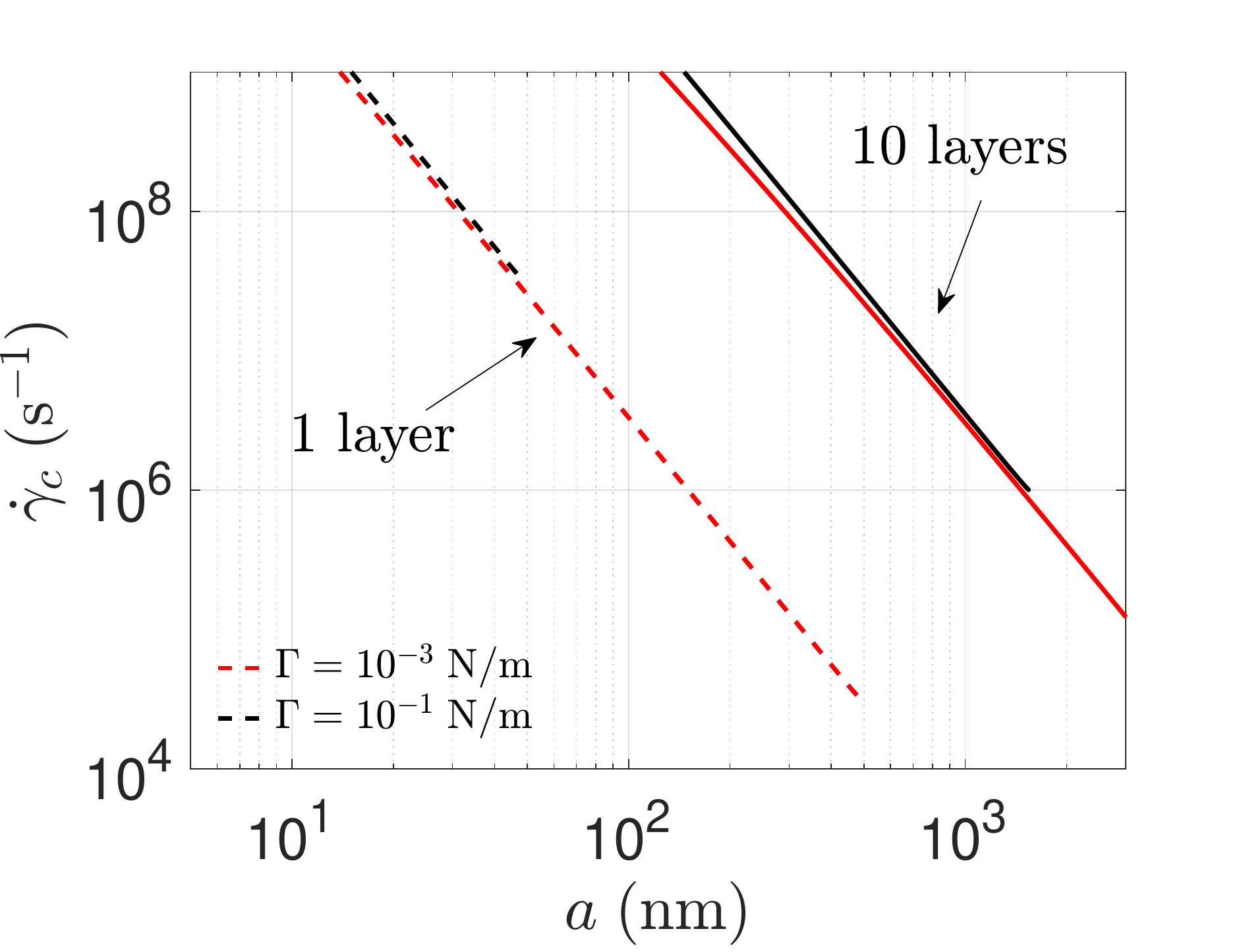}
\caption{Dimensional critical shear rate as a function of the size of the flap, for three different values of adhesion energy and two values of thickness (dashed lines for 1 layer, continuous lines for 10 layers). The value of the bending energy has been calculated as $D=D_0 N^3$ where $N$ is the number of layers and $D_0=20\;\mathrm{eV}$  \citep{sen2010tearing}. The value of the fluid viscosity is $\mu=10^{-3}\;\mathrm{Pa\;s}$.}
\label{fgr:gamma_a}
\end{figure}

A key result of our analysis is that the dependence of the hydrodynamic load on the opening of the flap can dramatically change the magnitude of the critical shear rate (see Fig. \ref{fgr:gamma_a}). We have identified a transition that occurs for values of $\Ghat$ in the range $10^{-5}-10^{-4}$. For  $\Ghat$ much smaller than this range of values, the constant load assumption holds and $\dot\gamma_c$ follows a power-law with an exponent $1/2$. For larger values of $\Ghat$,  $\dot\gamma_c$ follows Eq. (\ref{eqn:regime2}), which displays a weak dependence on adhesion. In this regime the critical shear rate $\dot\gamma_c$ is  much smaller than what predicted by a constant load assumption. This prediction is the manifestation of a self-reinforcing hydrodynamic effect: as the crack  propagates, the total pressure force on  the flap  increases  both because the length of the crack increases and because $\theta$ increases; the combination of these two effects increases the total force on the flap to a larger extent than if the pressure was considered independent of the wedge opening angle, producing large changes in flap curvature.  Interestingly, Fig. \ref{fgr:gamma_a} shows that our theory can predict relatively low values of the  critical shear rate of the order of $\sim 10^{5}s^{-1}$, close to those observed experimentally \citep{paton2014scalable}, even without assuming reductions in the adhesion energy  by several orders of magnitude when using specialised solvents (as instead assumed in the model of Ref. \citep{paton2014scalable}).

Unless the value of $\Ghat$ is truly infinitesimal, the error one would incur in by ignoring the transition we have discovered can be large. For example, from Fig.  \ref{fgr:shearrate12} we can see that   $\ghat_c/(8/q_1)$ is in the range $0.6-0.8$ when  $\Ghat = 0.1$.   The constant load solution  would give  a critical shear rate for exfoliation  one order of magnitude larger ($\ghat_c\cong8.94$ from Eq. (\ref{eq:G_cant})). In a practical liquid-exfoliation process, this difference would translate in drastically different processing conditions. In a rotating mixer for liquid-phase exfoliation, the average shear rate can be related to the mixer power $P$ and the liquid volume $V$ through $\dot{\gamma} \simeq \sqrt{P/(V \mu)}$ \citep{varrla2014}. Because of the scaling $P \propto \dot{\gamma}^2$, assuming $\Ghat = 0.1$ the constant load prediction would thus overestimate the mixing power by a factor of approximately 100. 

Expressions (\ref{eqn:regime1}) and (\ref{eqn:regime2}) suggest that to reduce the critical shear rate for exfoliation ( thus mitigating the possibility of fragmenting or causing mechanical damage to the exfoliated sheets)  one has to simultaneously reduce $\Gamma$ and increase $a$. The adhesion energy can be reduced by changing the solvent. However,  it has been reported that the dominant effect of adopting a good solvent is mostly to prevent reaggregation after exfoliation has taken place \citep{backes2016}, so it is not clear that good solvents can be designed that can change the critical shear rate by orders of magnitude.  Given the strong dependence on $a$ suggested by expressions (\ref{eqn:regime1}) and (\ref{eqn:regime2}), increasing  $a$ artificially could be a good strategy  to reduce the critical shear rate. This might be achieved by triggering a chemical reaction inside the layers  to enlarge pre-existing cracks \citep{martin1964}), exploiting electrostatic charge  \citep{cullen2017} or electrochemical effects \citep{shinde2016}.   Increasing $\mu$ also reduces the critical shear rate, but the overall stress level to which each particle is subject depends on the product $\mu \dot{\gamma}$.  Thus an increase in $\mu$ for a fixed $\dot{\gamma}$ may not be a solution if one wants to achieve a ``gentler'' exfoliation (using large viscosity fluids may still be beneficial because the reaggregation kinetics is slowed down \citep{santagiuliana2018}). 

We have assumed that an initial flaw is present. The fluid dynamics analysis reveal that, for a particle aligned with the streamline and for perfectly aligned edge layers (no shift between the layers), the direction of the load is such that, in the idealised situation, initiation of peeling starting from $a=0$ would be impossible. The direction of the load we find in our simulation is consistent with the result of Singh et al. for a disk aligned in a shear flow \citep{singh2014rotational}.   In a practical setting,  a small finite opening force would be present even in the case $a=0$, in instants in which the particle is inclined with respect to the flow direction or the edges of the particle are not perfectly aligned.

In the current work, we have focused on the range of moderately stiff flaps for which the wedge angle is smaller than $\pi/2$. Future work will explore larger values of the wedge angle. In this case, two aspects should be considered. First, for $\theta > \pi/2$ the pressure on the flap will start decreasing with increasing angle. Second, as the flap starts aligning with the long axis of the microparticle  tangential forces due to viscous shear stress will become dominant over pressure. These hydrodynamic features will yield new regimes of exfoliation, possibly extending the curve of Fig. \ref{fgr:shearrate12} to larger values of the non-dimensional adhesion energy. The analysis of these regimes will produce a more complete picture of the micromechanics of the exfoliation process.

\section*{Acknowledgements}
L. B. and G. S. acknowledge financial support from the EU through ERC Grant FlexNanoFlow (n. 715475) and Marie Curie Career Integration Grant FlowMat (n.618335). E. B. acknowledges funding by JSPS KAKENHI Grant Number JP18K18065. N. M. P. is
supported by the European Commission under the Graphene
Flagship Core2 No. 785219 (WP14 Composites), FET
Proactive Neurofibres grant No. 732344 and ARS01-01384-PROSCAN grant as well as by the Italian
Ministry of Education, University and Research (MIUR) under
the Departments of Excellence grant L.232/2016. 

\section*{Author contributions}
G. S. and  L. B. designed the research, developed the solid and fluid mechanics models, wrote the manuscript and analysed the results. N. M. P. designed the research, analysed the results and contributed to the development of the solid mechanics model. E. B. 
developed the solid mechanics model, analysed the results and contributed to writing the manuscript. All authors have approved the final version of the manuscript.

\section*{Competing interests statement}
The authors have no competing interests to declare.

\appendix

\section{Mathematical model for small displacements}
\label{sec:appendixA}

Denoting non-dimensional variables with a ``hat'' symbol (using $a$ and $D$ as repeating variables) the coupled equations for the small-displacement model are
\begin{subequations}
\begin{align}
\dfrac{\de^5 \widehat{w}_1}{\de \widehat{x}^5} = 0 \qquad 0 \leq \widehat{x} \leq 1   \\
\dfrac{\de^4 \widehat{w}_2}{\de \widehat{x}^4} + 4\chi^4\, \widehat{w}_2 = 0 \qquad 1 \leq \widehat{x}  \\
\dfrac{\de^4 \widehat{w}_1}{\de \widehat{x}^4}(0) - q_1\, \widehat{w}_1(0)\, \ghat = q_0 \, \ghat \\
\dfrac{\de^3 \widehat{w}_1}{\de \widehat{x}^3}(0)- f_1\, \widehat{w}_1(0)\,  \ghat = f_0\,  \ghat \\
\dfrac{\de^2 \widehat{w}_1}{\de \widehat{x}^2} (0)= 0 \\
\widehat{w}_1(1) = \widehat{w}_2(1) \label{eq:cont1} \\
\dfrac{\de \widehat{w}_1}{\de \widehat{x}} (1)= \dfrac{\de \widehat{w}_2}{\de \widehat{x}} (1)\label{eq:cont2}\\
\dfrac{\de^2 \widehat{w}_1}{\de \widehat{x}^2} (1) = \dfrac{\de^2 \widehat{w}_2}{\de \widehat{x}^2} (1) \label{eq:cont3} \\
\dfrac{\de^3 \widehat{w}_1}{\de \widehat{x}^3} (1) =\dfrac{\de^3 \widehat{w}_2}{\de \widehat{x}^3} (1)\label{eq:cont4} 
\end{align}
\label{eq:Model_Adim_FINAL}
\end{subequations}
In non-dimensional units, Griffith's energy balance is given by
\begin{equation}
\widehat{G} = \Ghat 
\end{equation}
with 
\begin{equation}
\widehat{G} = 3\, \frac{\partial \widehat{U}}{\partial\ghat}\ghat + \dfrac{\partial \widehat{U}}{\partial \chi}\, \chi - \widehat{U}
\end{equation}
and 
\begin{equation}
\widehat{U} = \frac{1}{2}\int_0^1 \left( \widehat{w}_1'' \right)^2  \de \widehat{x} + \frac{1}{2} \int_1^{\infty} \left( \widehat{w}_2''\right)^2  \de \widehat{x}
\end{equation}

The small-displacement solutions for $\widehat{w}_1(\widehat{x})$ and $\widehat{w}_2(\widehat{x})$ are
\begin{align}
\widehat{w}_1=&\frac{\ghat}{6\ghat\chi\Big(\chi^3(8f_1+3q_1)+12\chi^2(2f_1+q_1)+6\chi(4f_1+3q_1)+12(f_1+q_1)\Big)-144\chi^4}\times\notag\\
\times&\Big[\ghat \widehat{x}(f_1q_0-f_0q_1)\Big(\chi^4(2\widehat{x}^3-3\widehat{x}^2+1)+6\chi^3(\widehat{x}^3-2\widehat{x}^2+1)+3\chi^2(2\widehat{x}^3-6\widehat{x}^2+5)\notag\\
&+3\chi(\widehat{x}^3-4\widehat{x}^2+8)+18\Big)-6\chi\;\Big(\chi^3(q_0\widehat{x}^4+4f_0\widehat{x}^3-4(q_0+3f_0)\widehat{x}+3q_0+8f_0)\notag\\
&+12\chi^2(q_0+2f_0)(1-\widehat{x})+6\chi(3q_0+4f_0-2(q_0+f_0)\widehat{x})+12(q_0+f_0)\Big)\Big]\\[5mm]
\widehat{w}_2=&\frac{\ghat e^{\chi-\chi \widehat{x}}}{2\ghat\chi^2\Big(\chi^3(8f_1+3q_1)+12\chi^2(2f_1+q_1)+6\chi(4f_1+3q_1)+12(f_1+q_1)\Big)-48\chi^5}\times\notag\\
\times&\Big[\ghat(f_1q_0-f_0q_1)\Big(\sin(\chi-\chi \widehat{x})(\chi^3-6\chi-6)+\cos(\chi-\chi \widehat{x})(\chi^3+5\chi^2+6\chi)\Big)\notag\\
&-12\chi^3(2f_0+q_0)\Big(\sin(\chi-\chi \widehat{x})+\cos(\chi-\chi \widehat{x})\Big)-24\chi^2\cos(\chi-\chi \widehat{x})(f_0+q_0)\Big]
\end{align}
The strain energy release rate is a rational function of polynomial functions in $\chi$ and $\ghat$
\begin{equation}
\widehat{G} = \frac{N_{\widehat{G}}(\chi,\ghat)}{D_{\widehat{G}}(\chi,\ghat)}
\end{equation}
which becomes
\begin{equation}
N_{\widehat{G}}(\chi,\ghat) = D_{\widehat{G}}(\chi,\ghat)\, \Ghat
\end{equation}
This function is a quintic polynomial in $\ghat$
\begin{equation}
c_5\, \ghat^5 + c_4\, \ghat^4 + c_3\, \ghat^3 + c_2\, \ghat\,^2 + c_1\, \ghat + c_0 = 0
\end{equation}
with coefficients
\begin{equation}
c_0=552960\; \Ghat \chi^{10}
\end{equation}
\begin{equation}
c_1=-69120\, \Ghat\,\chi^7\Big[4f_1(3+6\chi+6\chi^2+2\chi^3)+3q_1(4+6\chi+4\chi^2+\chi^3)\Big]
\end{equation}
\begin{align}
c_2=&2880\,\chi^4\Big[\Ghat\Big(4f_1(3+6\chi+6\chi^2+2\chi^3)+3q_1(4+6\chi+4\chi^2+\chi^3)\Big)^2\notag\\
&-24\,\chi^3\Big(f_0^2(1+6\chi+12\chi^2)+f_0q_0(2+9\chi+12\chi^2)+q_0^2(1+3 \chi + 3 \chi^2)\Big)\Big]
\end{align}
\begin{align}
c_3=&40\chi\Big[-\Ghat \Big(4f_1(3 + 6 \chi + 6 \chi^2 + 2 \chi^3) + 3 q_1(4 + 6 \chi + 4 \chi^2 + \chi^3)\Big)^3\notag\\
&+72\chi^4(f_1q_0-f_0q_1)\Big(f_0(-18 - 96 \chi - 35 \chi^2 + 72 \chi^3+ 42\chi^4)\notag\\
&+q_0(-18 - 48 \chi - 5\chi^2 + 42 \chi^3 + 21 \chi^4)\Big)\notag\\
&+36\chi^3\Big(2f_0q_0(4f_1(6+27\chi + 42 \chi^2 + 10 \chi^3 - 18 \chi^4 - 12 \chi^5)\notag\\
&+ 3q_1(8 + 36 \chi + 58 \chi^2 + 20 \chi^3 - 9 \chi^4 - 6 \chi^5) )\notag\\
    & + q_0^2 (-8 f_1 (-3 - 9 \chi - 9 \chi^2 + \chi^3 + 6 \chi^4 + 3 \chi^5) \notag\\
    &+ 3 q_1(8 + 24 \chi + 28 \chi^2 + 6 \chi^3 - 6 \chi^4 - 3 \chi^5) ) \notag\\
    &- 4 f0^2 (f_1 (-6 - 36 \chi - 84 \chi^2 - 40 \chi^3 + 24 \chi^4 + 24 \chi^5) \notag\\
    &+3 q_1(-2 - 12 \chi - 28 \chi^2 - 13 \chi^3 + 3 \chi^4 + 3 \chi^5)) \Big)\Big]
\end{align}
\begin{align}
c_4=&-24\, \chi^2\,\Big[2 (f_1q_0-f_0q_1) \chi (1 + \chi) (540 + 1320 \chi + 1578 \chi^2 + 1194 \chi^3 + 592 \chi^4 + 173 \chi^5 + 22 \chi^6)\notag\\
&+20 q_0 f_1 (-54 - 180 \chi - 123 \chi^2 + 144 \chi^3 + 273 \chi^4 + 176 \chi^5 + 54 \chi^6 + 6 \chi^7)\notag\\
&+15q_0q_1 (-72 - 228 \chi - 140 \chi^2 + 168 \chi^3 + 272 \chi^4 + 147 \chi^5 + 36 \chi^6 +3 \chi^7)\notag\\
&+20 f_0 f_1 (-54 - 324 \chi - 357 \chi^2 + 126 \chi^3 + 468 \chi^4 + 338 \chi^5 + 108 \chi^6 + 12 \chi^7)\notag\\
&+15f_0q_1(-72 - 420 \chi - 404 \chi^2 + 180 \chi^3 + 476 \chi^4 + 283 \chi^5 + 72 \chi^6 + 6 \chi^7) \Big](f_1q_0-f_0q_1)
\end{align}
\begin{align}
c_5=&\Big[40 f_1 (1 + \chi) (6 + 6 \chi + 6 \chi^2 + 4 \chi^3 + \chi^4) (54 + 132 \chi + 141 \chi^2 + 84 \chi^3 + 28 \chi^4 + 4 \chi^5)\notag\\
 &+3q_1 (4320 + 18120 \chi + 36576 \chi^2 + 47796 \chi^3 + 45032 \chi^4 + 31698 \chi^5 \notag\\
 &+ 16636 \chi^6 + 6341 \chi^7 + 1662 \chi^8 + 268 \chi^9 + 20 \chi^{10})\Big](f_1q_0-f_0q_1)^2
\end{align}
In the simpler case considered in Section \ref{sec:res_solid} (distributed load only, independent on the angle), the displacements reduces to
\begin{equation}
\widehat{w}_1(\widehat{x})=q_0\ghat\frac{\chi^3(\widehat{x}^4-4\widehat{x}+3)-12\chi^2(\widehat{x}-1)-6\chi(2\widehat{x}-3)+12}{24\chi^3}
\end{equation}
\begin{equation}
\widehat{w}_2(\widehat{x})=q_0\ghat\mathrm{e}^{\chi(1-\widehat{x})}\frac{2\cos(\chi(\widehat{x}-1))+\chi\cos(\chi(\widehat{x}-1))+\chi\sin(\chi(\widehat{x}-1))}{4\chi^3}
\end{equation}
from which the relation between $\ghat$  and $\Ghat$ has been calculated
\begin{equation}\label{eq:app_Gsoft}
\ghat=\frac{2\sqrt{2}}{q_0}\Ghat^{\frac{1}{2}}\left(\frac{\chi}{1+\chi}\right)^{3/2}
\end{equation}

If the angle independent edge load $\tilde{F}=f_0$ is applied together with the distributed load $q=q_0$ the displacements become
\begin{align}
\widehat{w}_1(\widehat{x})=&\frac{q_0\ghat}{24\chi^3}\left(\chi^3(\widehat{x}^4-4\widehat{x}+3)-12\chi^2(\widehat{x}-1)-6\chi(2\widehat{x}-3)+12\right)\notag\\
&+\frac{f_0\ghat}{4\chi^3}\left(\chi^3(\widehat{x}^3-3\widehat{x}+2)-4\chi^2(\widehat{x}-1)-3\chi(\widehat{x}-2)+3\right)
\end{align}
\begin{align}
\widehat{w}_2(\widehat{x})=&\frac{q_0\ghat\mathrm{e}^{\chi(1-\widehat{x})}}{4\chi^3}\left(2\cos(\chi(\widehat{x}-1))+\chi\cos(\chi(\widehat{x}-1))+\chi\sin(\chi(\widehat{x}-1))\right)\notag\\
&\frac{f_0\ghat\mathrm{e}^{\chi(1-\widehat{x})}}{2\chi^3}\left(\cos(\chi(\widehat{x}-1))+\chi\cos(\chi(\widehat{x}-1))+\chi\sin(\chi(\widehat{x}-1))\right)
\end{align}
The relation between $\ghat$ and $\Ghat$ shows the same dependence on $\Ghat$ as Eq. (\ref{eq:app_Gsoft}), with a prefactor that depends also on $f_0$
\begin{equation}\label{eq:app_Gsoft_f0}
\ghat=\frac{2\sqrt{2}}{q_0}\frac{\chi^{3/2}}{\left[(1+\chi)^3+f_0/q_0(1+\chi^2)(2+5\chi)+f_0^2/(3q_0^2)(3+18\chi+36\chi^2+20\chi^3)\right]^{1/2}}\Ghat^{\frac{1}{2}}
\end{equation}

\subsection{Infinitely stiff foundation: cantilever beam ($\chi\rightarrow\infty $)}\label{sec:appendixA1}
If the foundation is considered as infinitely stiff, the beam $\mathcal{B}_1$ can be seen as a clamped beam.
The solution for the displacement if both the distributed load and the edge load are applied is
\begin{equation}
\widehat{w}_1(\widehat{x})=\frac{\ghat(1-\widehat{x})^2\Big(\ghat(f_1q_0-f_0q_1)(2\widehat{x}^2+\widehat{x})-6q_0(\widehat{x}^2+2\widehat{x}+3)-24f_0(\widehat{x}+2)\Big)}{6(\ghat(8f_1+3q_1)-24)}
\end{equation}
The strain energy obtained from the displacement is
\begin{equation}\label{eq:U_qf}
\widehat{U}=\frac{\ghat\,^2[(144 + f_1\ghat(-6 + f_1\ghat))q_0^2 + 2f_0 q_0(360 + \ghat (3 - f_1\ghat) q_1) +f_0^2(960 + \ghat\,^2q_1^2)]}{10 (-24 + 8 f_1 \ghat + 3 q_1\ghat)^2}
\end{equation}
Again, the Griffith's energy balance can be written as
\begin{equation}\label{eq:quintic}
c_5\, \ghat^5 + c_4\, \ghat^4 + c_3\, \ghat^3 + c_2\, \ghat\,^2 + c_1\, \ghat + c_0 = 0
\end{equation}
with coefficients  
\begin{equation}
c_0=138240 \Ghat
\end{equation}
\begin{equation}
c_1=-17280\Ghat(8 f_1 + 3 q_1)
\end{equation}
\begin{equation}
c_2=720\Big[\Ghat(8 f_1 + 3 q_1)^2-8 (20 f_0^2 + 15 f_0 q_0 + 3 q_0^2)\Big]
\end{equation}
\begin{equation}
c_3=-10\Ghat (8 f_1 + 3 q_1)^3-48 (160 f_0^2 f_1 + 120 f_0 f_1 q_0 + 60 f_0^2 q_1 + 69 f_0 q_0 q_1 + 9 q_0^2 q_1)
\end{equation}
\begin{equation}
c_4=-12 (f_1 q_0 - f_0 q_1) (30 f_1 q_0 - 22 f_0 q_1 + 3 q_0 q_1)
\end{equation}
\begin{equation}
c_5=5 (8 f_1 + 3 q_1) (f_1 q_0 - f_0 q_1)^2
\end{equation}

If the applied load consists in a distributed and an edge load that do not depend on the angle, the solution simplifies to
\begin{equation}
\widehat{w}_1(\widehat{x})=\frac{\ghat}{24}(\widehat{x}-1)^2 \left( q_0(\widehat{x}^2+2\widehat{x}+3)+4 f_0(\widehat{x}+2)\right).
\end{equation}
The corresponding solution to the Griffith's energy balance is 
\begin{equation}
\ghat=\frac{2\sqrt{2}}{q_0}\frac{1}{\sqrt{1+5f_0/q_0+20/3(f_0/q_0)^2}}\Ghat^{\frac{1}{2}}.
\end{equation}

If an angle-dependent, distributed load is applied, the equilibrium shape is 
\begin{equation}
\widehat{w}_1(\widehat{x})=\frac{\ghat q_0  }{3 (8 - \ghat q_1)}(\widehat{x}-1)^2(\widehat{x}^2+2\widehat{x}+3)
\end{equation}
and the Griffith's energy balance (\ref{eq:quintic}) simplifies to a cubic polynomial in $\ghat$
\begin{equation}
512\;\Ghat - 192 \;\Ghat q_1 \ghat  + 8\ghat\,^2 (-8\; q_0^2 + 3 \;\Ghat q_1^2) + \ghat^3 (-\frac{8}{5}\; q_0^2 q_1 - \;\Ghat q_1^3) = 0.
\end{equation}

If the distributed load is independent on the angle, the classic solution for the cantilever beam under uniform load is recovered
\begin{equation}
\widehat{w}_1(\widehat{x})=\frac{\ghat q_0}{24}(\widehat{x}-1)^2(\widehat{x}^2+2\widehat{x}+3)
\end{equation}
with the Griffith's energy balance giving
\begin{equation}
\frac{\mu\dot\gamma a^3}{D}=\frac{2\sqrt{2}}{q_0}\left(\frac{\Gamma a^2}{D}\right)^{\frac{1}{2}}.
\end{equation}




\bibliographystyle{unsrt}
\bibliography{biblio}
\end{document}